\documentclass[pdflatex,sn-mathphys-num]{sn-jnl}

\usepackage{subcaption}%
\usepackage{url}%
\usepackage{graphicx}%
\usepackage{multirow}%
\usepackage{amsmath,amssymb,amsfonts}%
\usepackage{amsthm}%
\usepackage{mathrsfs}%
\usepackage{textcomp}%
\usepackage{manyfoot}%
\usepackage{booktabs}%
\usepackage{algorithm}%
\usepackage{algorithmicx}%
\usepackage{algpseudocode}%
\usepackage{listings}%
\usepackage{forest}%
\usepackage[svgnames]{xcolor} 
\usepackage{float}%
\usepackage{adjustbox}%
\usepackage{soul}%

\theoremstyle{thmstyleone}%
\theoremstyle{thmstyletwo}%

\theoremstyle{thmstylethree}%

\begin{document}

\title[Article Title]{Large Language Models for Software Engineering Diagrams: A Systematic Review of UML and ER modelling}


\author[1]{\fnm{Mojdeh} \sur{Rahmanian}}\email{m.rahmanian@napier.ac.uk}

\author*[1]{\fnm{Ashkan} \sur{Sami}}\email{a.sami@napier.ac.uk}

\author[1]{\fnm{Yanchao} \sur{Yu}}\email{y.yu@napier.ac.uk}

\affil*[1]{\orgdiv{School of Computing, Engineering and The Built Environment}, \orgname{Edinburgh Napier University}, \orgaddress{\street{Colinton}, \city{Edinburgh}, \postcode{EH10 5DT}, \state{State}, \country{UK}}}


\abstract{Large language models (LLMs) are increasingly applied to diagram-based software and data modelling. Among various modelling notations, UML and entity-relationship (ER) diagrams are the most widely adopted for software modelling and data modelling, respectively.
Recent literature has investigated various applications of LLMs in diagram modelling; however, their effectiveness and limitations have not been extensively discussed. This systematic literature review analyses 64 studies published between 2023 and 2025, examining diagram coverage, modelling tasks, technical approaches, evaluation practices, and limitations. 
Our findings reveal significant concentration patterns and gaps. UML-based software modelling strongly dominates, with class diagrams receiving the most attention whilst behavioural diagrams and data modelling remain underrepresented. Diagram construction from natural language is the primary focus, with limited work on transformation, quality assurance, and consistency checking. GPT-based models are heavily prevalent, raising concerns about reproducibility and vendor dependence. Evaluation practices are heterogeneous, employing diverse metrics and custom datasets with limited benchmark reuse and inconsistent reporting of robustness and statistical significance. Common limitations include semantic inaccuracies, hallucinated diagram elements, sensitivity to prompt formulation, and reproducibility constraints.

This survey provides the first systematic synthesis of LLM-based diagram modelling research, highlighting needs for standardised benchmarks, stronger evaluation protocols, broader diagram coverage, and techniques for improving semantic reliability and multi-view consistency.
}

\keywords{Large language models; Software modelling; UML;ER diagram; Systematic literature review;}



\maketitle
\section{Introduction}\label{sec: Intro} 

Software engineering (SE) has long relied on models and diagrams as key tools for understanding, communicating, and developing complex systems. This idea is highlighted in the foundational work on model-driven engineering \cite{booch_unified_2005, schmidt_guest_2006}. Notations such as UML and domain-specific architecture frameworks provide a common languages that captures the structure and behavior of systems \cite{sysml_systems_2006}. These diagrammatic notations act as a vital link between the informal needs of stakeholders and the executable implementations, a topic that has been thoroughly explored in requirements engineering and modelling research \cite{jackson_problem_2000, sommerville_software_2011}

Diagrams play a crucial role in capturing the structural elements like classes, components, and deployment nodes that are essential for architectural and design reasoning
\cite{clements_documenting_2003, fowler_uml_2018}. They also illustrate behavioral aspects through tools like sequence diagrams, state machines, and activity models \cite{rumbaugh_unified_2006}. However, despite their significance, the process of creating and maintaining high-quality diagrams can be quite labor-intensive, prone to errors, and often not practiced enough in industrial settings, as highlighted by various empirical studies on modelling practices  \cite{petre_uml_2013,hutchinson_empirical_2011}.
At the same time, recent advances in LLMs have transformed natural language processing and, increasingly, the field of software engineering. For example, OpenAI Codex, which was trained on public code repositories, has proven it can generate valid code from natural language specifications \cite{chen_evaluating_2021}. Additionally, newer models like GPT-4 have demonstrated impressive capabilities across multiple domains, including coding, reasoning, and cross-domain tasks.  This has led to an increasing confidence in the potential of LLMs to assist with complex software tasks, such as generating code, assessing its correctness, and even engaging in creative problem-solving \cite{tambon_bugs_2025, mastropaolo_robustness_2023}.

These advancements raise an important question: how can large language models support or automate software modelling activities? 
Recent studies have explored the use of LLMs in various modelling tasks, including generating artifact generation, transformation, and analysis. 
However, these studies differ significantly in how they approach tasks, their evaluation methods, and how they fit into existing model-driven engineering workflows. Additionally, the literature is spread across various venues and research communities, making it challenging to get a cohesive understanding.The current landscape of research is missing a unified perspective on its scope, methods, and limitations. When it comes to working with diagrams, the challenges we face are quite different from the more straightforward tasks in software engineering, like code completion. 
Diagrammatic notations are shaped by metamodels, well-formedness rules, and strict syntactic and semantic constraints, as noted in MDE research \cite{kleppe_mda_2003,selic_what_2012}. For effective modelling, it’s crucial to maintain consistency across different views, ensuring that structural, behavioral, and architectural diagrams remain aligned as systems evolve \cite{spanoudakis_inconsistency_2001,nuseibeh_leveraging_2002}.

Using LLMs to create diagrams from natural language can sometimes lead to results that look convincing but actually break grammatical or meaning rules. This can result in "hallucinated" elements, mismatched relationships, or the wrong application of domain frameworks. 
However, LLMs also offer strong capabilities for semantic interpretation, summarization, and explanation, which may overcome some past challenges faced by rule-based or information-retrieval methods in transforming requirements into models
\cite{landhauser_requirements_2014,hutchison_automatic_2005,gropler_nlp-based_2021}.

Although recent studies report encouraging results, the literature lacks a consolidated view of what large language models currently achieve in the context of software and data diagram modelling. The scope of supported diagram types, the range of modelling tasks, the effectiveness of different technical configurations, and the robustness of evaluation practices remain unevenly documented.
To fill this gap, we conduct a systematic review of research on the use of large language models for UML and entity–relationship (ER) diagram modelling. These notations represent the two dominant modelling concerns in software engineering: software modelling and data modelling. UML is the ISO/IEC 19505 standard for modelling software systems and is the most widely studied diagrammatic notation in the field of model-driven engineering \cite{marcen2024systematic}. 
 Entity-relationship modelling, introduced by Chen \cite{chen1976entity} and widely adopted in database design, focuses specifically on conceptual data representation. This distinction between system-level and data-level modelling is well-established in software engineering practice \cite{sommerville2011software}, where UML-based design and ER-based database modelling are taught and applied as complementary modelling skills.
ER diagrams share key structural modelling features with UML class diagrams, such as entities, attributes, relationships, and cardinality constraints.

We acknowledge that software engineering practice employs additional diagrammatic notation, including BPMN and SysML. However, these notation are governed by different metamodels, serve different engineering contexts, and are primarily addressed within separate research communities. By concentrating on UML and ER diagrams, this survey keeps a clear and focused direction while highlighting the key modelling notations prevalent in the current literature.

The review is guided by the following research questions:

\begin{itemize}
\item
  \textbf{RQ1:} Which types of UML and entity-relationship diagrams are addressed in LLM-based research?
\item
  \textbf{RQ2:} What modelling tasks are supported by LLM-based approaches, and how are these tasks formulated?
\item
  \textbf{RQ3:} What LLM models, prompting strategies, and system architectures are employed?
\item
  \textbf{RQ4:} How are these approaches evaluated, and what datasets and evaluation strategies are used?
\item
  \textbf{RQ5:} What limitations are reported in the current body of work?
\end{itemize}

These questions guided our search, selection, and data extraction
processes.

The main contributions of this survey can be summarized as follows:

\begin{enumerate}
\item A systematic and focused review of research on the use of large language models in software modelling, providing a clear overview of current work within dominant modelling notations.
\item A structured characterization of the diagram types and modelling tasks addressed in the literature, highlighting concentration patterns and areas with limited empirical evidence.
\item An analysis of prevailing strategies for integrating large language models into modelling workflows, including common prompting approaches and system configurations.
\item An overview of evaluation practices, examining the metrics, datasets, and experimental designs used to assess performance and clarifying the current empirical basis of the field.
\item  An analysis of common technical and methodological limitations reported across studies, highlighting current capabilities and constraints.
\end{enumerate}

The remainder of this paper is organized as follows. Section 2 describes the review protocol and data extraction procedure. Section 3 presents the results of the survey, including diagram types, modelling tasks, technical integration strategies, evaluation practices, and reported limitations. Section 4 discusses the key observations and implications of these findings. Section 5 presents a broader discussion of the results. Section 6 outlines directions for future research. Section 7 discusses threats to validity, and Section 8 concludes the paper.

\section{Related works}\label{sec2:RelatedWork}

This section places our current survey within the wider context of existing reviews and surveys. We have organized the discussion around four key themes that together outline the scope and significance of our work.

\textbf{LLMs across the Software Engineering Lifecycle:} Several systematic literature reviews have explored how LLMs are being used throughout the software engineering lifecycle. Hou et al. \cite{hou2024large}   present one of the most comprehensive surveys available, looking at various software engineering tasks like code generation, testing, maintenance, and repair. Zhang et al. \cite{zhang2023survey}  focus their review specifically on code-related tasks across five phases of software engineering. Fan et al. \cite{fan2023large}   take a forward-thinking approach, discussing code generation, testing, and maintenance, and they notably point out that many practitioners are still hesitant to depend on LLMs for more advanced design objectives. Liu et al. \cite{liu2024large} broaden the discussion by examining LLM-based agents in software engineering, highlighting how LLMs, when paired with external tools, memory, and planning abilities, can tackle more intricate and autonomous software engineering tasks. A common theme in these surveys is that research on LLMs in software engineering is primarily focused on code-centric activities. Software design and modelling are recognized as significant but receive only limited attention, and none of these works includes a dedicated analysis of diagram generation, the modelling tasks it involves, or the evaluation criteria it requires.

\textbf{Model-Driven Engineering and Machine Learning:}
A separate but closely related line of survey research that looks at how model-driven engineering (MDE) intersects with machine learning (ML). In their systematic review, Marcén et al. \cite{marcen2024systematic} dive into studies where ML techniques, like neural networks, decision trees, and support vector machines are applied to MDE challenges such as feature location, model transformation, and recovering traceability links. However, it's worth noting that much of their work predates the rise of generative large language models. The reverse direction, Naveed et al. \cite{naveed2024model}  and Rädler et al. \cite{radler2025bridging}  have explored how MDE techniques can aid in AI development, but they found that current methods are still quite immature and lack robust tools. More relevant to our current discussion, Di Rocco et al. \cite{di2025use}  investigate how LLMs can be applied within MDE, focusing on areas like model synthesis and repository classification. They even suggest prompting strategies that are tailored to different modelling contexts, whether you're working with raw text or templates. Meanwhile, Burgueño et al. \cite{burgueno2025automation}  outline a roadmap for automating MDE, recognizing the transformative potential of LLMs, but they also point out that generating visual diagrams from natural language is still a bit scattered in the literature. Despite the increasing interest in the relationship between LLMs and MDE, none of these surveys have thoroughly tackled the end-to-end generation of formally correct UML or ER diagrams using large language models.

\textbf{LLMs for Requirements Engineering:}
Requirements engineering is the initial phase of software development where modelling activities naturally come into play, and several recent surveys have looked into the role of LLMs in this area. Hemmat et al. \cite{hemmat2025research}  explore how LLMs are utilized across various RE tasks, highlighting outputs like UML diagrams, while pointing out that the transformation of requirements into models is still relatively underexplored compared to text-focused RE activities. Zadenoori et al. \cite{zadenoori2025large}  take a broader approach and specifically categorize requirements modelling as a task of interest, although it still takes a backseat to elicitation and validation. Their analysis also shows that zero-shot and few-shot prompting are the most common methods used, while more advanced techniques like chain-of-thought reasoning and retrieval-augmented generation are rarely applied in modelling scenarios. Cheng et al. \cite{cheng2026generative}   further expand their review, noting that issues like reproducibility, hallucinations, and interpretability are closely linked challenges found throughout the RE–LLM literature. Across all three surveys, generating diagrams from requirements is recognized as a capability, but it is consistently seen as an area that needs more focused research and stronger evaluation frameworks.

\textbf{Generative AI for Software Architecture:}
Recent surveys have started to examine how generative AI is influencing software architecture, which is closely tied to software modelling. Esposito et al. \cite{esposito2025generative}  carried out a comprehensive literature review and discovered that the main focus of current research is on architectural decision support and architectural reconstruction. They noted that techniques like few-shot prompting and retrieval-augmented generation are leading the way, while UML diagram generation seems to be only a minor player. On the other hand, Schmid et al. \cite{schmid2025software}  zoom in on the overlap between software architecture and LLMs, specifically leaving out UML models like class and activity diagrams from their analysis. This confirms that diagram-level modelling isn't a priority for the software architecture community right now. Both surveys highlight a significant challenge: the lack of robust evaluation frameworks and architecture-specific benchmarks, which echoes the evaluation gaps pointed out in the current work.

\textbf{Positioning of This Survey}
While the general LLM-for-SE surveys \cite{hou2024large,fan2023large,zhang2023survey,liu2024large} cover the whole lifecycle, they tend to miss the mark on diagramming. Meanwhile, MDE surveys \cite{marcen2024systematic,di2025use,burgueno2025automation,naveed2024model,radler2025bridging} tackle modelling but do not fully address LLM-based tasks. To this point, no prior survey has systematically and thoroughly explored how large language models are utilized for generating, transforming, evaluating, and understanding software engineering diagrams. Our survey aims to fill this gap by focusing on various types of diagrams, particularly UML and ER, the specific modelling tasks that LLMs can assist with, the technical setups and prompting strategies employed, the evaluation methods used, and the limitations noted in the existing literature.

\section{Survey Methodology}\label{sec3:Methodology}

To systematically investigate the role of LLMs in software engineering diagramming, we conducted a systematic literature review following widely adopted guidelines for evidence-based software engineering 
\cite{keele_guidelines_2007}. The objective of this survey is to identify, analyse, and synthesize peer-reviewed studies that explicitly apply LLM-based techniques, such as GPT-style models or other generative AI systems, to
tasks involving UML or related diagrammatic modelling notations.

\subsection{Paper Collection Strategy}\label{subsec2:PaCo}

\begin{figure}[h]
\centering
\includegraphics[width=0.9\textwidth]{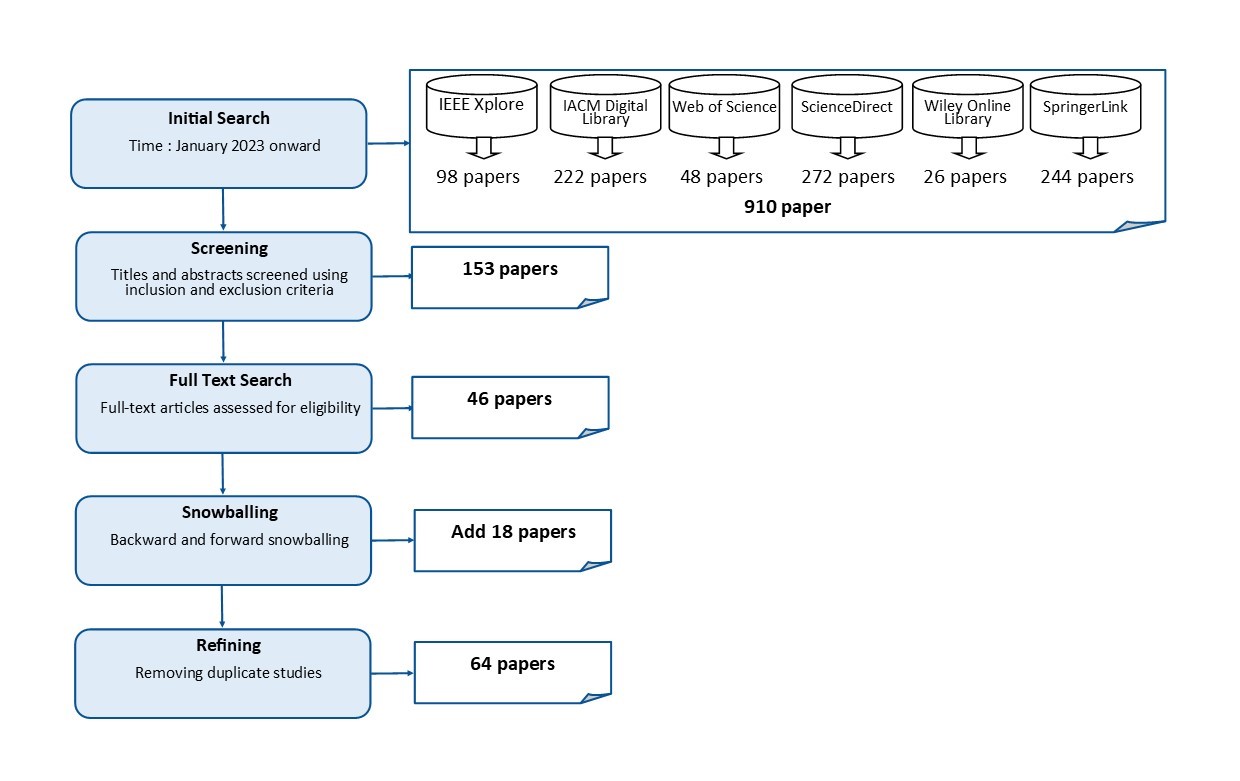}
\caption{Paper collection and selection process.}\label{fig:processOverview}
\end{figure}

To collect relevant studies, we employed a systematic paper collection strategy that combines keyword-based database search with manual snowballing. The search process was designed to capture research at the intersection of large language models and software engineering diagrams. The overall paper collection and selection process, is depicted in Fig.~\ref{fig:processOverview}.

We used the following search string consistently across all selected
digital libraries: \emph{("large language model" OR LLM OR "GPT" OR
"ChatGPT" OR "generative AI" OR "Gen AI") AND ("UML" OR "class diagram"
OR "sequence diagram" OR "use case diagram" OR "activity diagram" OR "ER
diagram" OR "entity relationship model").} This search string targets
studies that explicitly combine LLM-based or generative AI techniques
with diagrammatic modelling notations commonly used in software
engineering.

The search was conducted across major scientific digital libraries that
serve as primary publication venues for software engineering research,
including IEEE Xplore, the ACM Digital Library, Web of Science, Elsevier
ScienceDirect, Wiley Online Library, and SpringerLink. These digital
libraries were selected to ensure broad coverage of peer-reviewed journal articles and conference proceedings in software engineering and
related fields.

The search was limited to studies published from January 2023 onward. This time frame was chosen because the widespread adoption of large language models in software engineering began after the public release of GPT-3.5 and GPT-4, which marked a qualitative shift in the practical applicability of LLMs to complex engineering tasks. Earlier work primarily focused on pre-LLM techniques and is therefore outside the scope of this survey.

The collected papers were subsequently filtered based on predefined inclusion and exclusion criteria, which are summarized in Table~\ref{tab:criteria}. To ensure the reliability and consistency of the study selection process, two authors independently performed the screening of papers based on the predefined inclusion and exclusion criteria. This process was applied during the title and abstract screening stages. Disagreements between the reviewers were resolved through discussion until consensus was reached. To quantify inter-rater agreement, Cohen’s Kappa coefficient was computed. The resulting Kappa value was 0.738, indicating substantial agreement. These results demonstrate a strong level of consistency in the study selection process and support the reliability of the inclusion decisions.

\begin{table}[h]
\caption{Inclusion and Exclusion Criteria}\label{tab:criteria}
\begin{tabular}{@{}lp{0.85\textwidth}@{}}
\toprule
\multicolumn{2}{@{}l@{}}{\textbf{Inclusion Criteria}} \\
\midrule
(1) & The paper applies large language models or generative AI as a core component of the proposed approach. \\
(2) & The paper addresses software engineering tasks involving diagrammatic modelling artifacts, including UML diagrams, ER diagrams, or architectural and domain-specific modelling notations.\\
(3) & The paper treats diagrams as primary artifacts, either as inputs, outputs, or analysis targets (e.g., generation, extraction, evaluation, or support tasks). \\
(4) & The paper presents an implemented method or tool and includes an evaluation related to UML or ER diagram–centric tasks. \\
\midrule
\multicolumn{2}{@{}l@{}}{\textbf{Exclusion Criteria}} \\
\midrule
(1) & The paper does not involve large language models or generative AI techniques. \\
(2) & The paper is purely conceptual or does not include any evaluation. \\
(3) & The paper focuses on diagram types outside UML or ER, or uses diagrams only for illustration. \\
(4) & The paper is not written in English. \\
(5) & The paper is published only as a preprint (e.g., arXiv) and has not appeared in a peer-reviewed venue. \\
\botrule
\end{tabular}
\end{table}

\subsection{Snowballing}\label{subsec2:SB}

To enhance the completeness of the survey and reduce the risk of missing
relevant studies, we adopted both backward and forward snowballing
techniques following established survey practices \cite{chen_fairness_2024}. Backward
snowballing was performed by examining the reference lists of all papers
selected after full-text screening, while forward snowballing was
conducted using Google Scholar to identify studies citing the selected
papers.

The snowballing process was carried out iteratively until January 1, 2026, and continued until no new relevant papers were
identified. This process enabled the inclusion of additional relevant
studies that were not captured by the initial database search.

\section{Results}\label{subsec2:Result}
\subsection{Overview of Included Studies}\label{subsec2:inSt}

Our final dataset comprises 46 primary studies identified through database search and screening, complemented by 18 additional studies identified through backward and forward snowballing, yielding a total of 64 studies. These works represent the current state of research on applying LLMs to software engineering diagram–centric tasks.

\begin{figure}[h]
\centering
\includegraphics[width=0.7\textwidth]{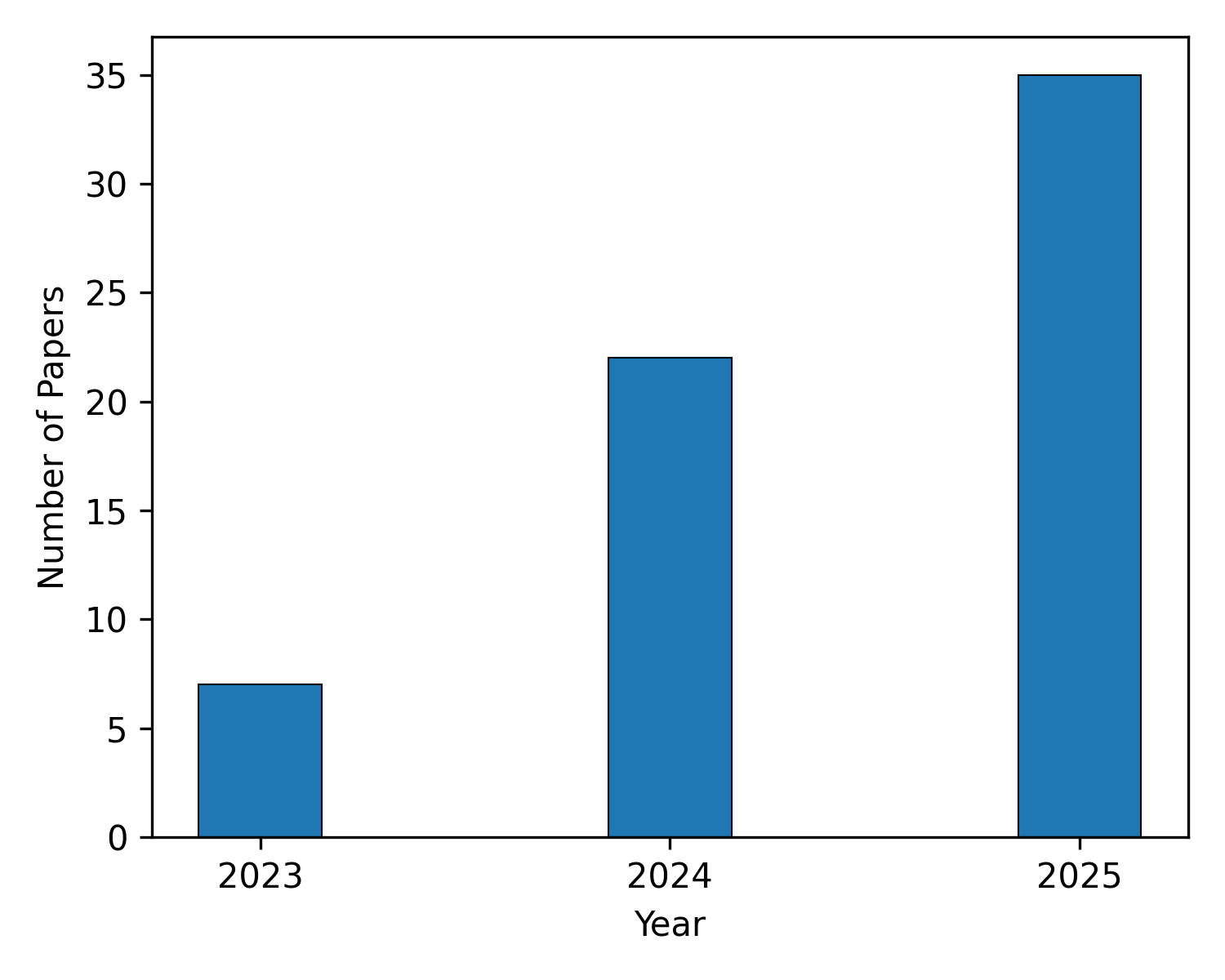}
\caption{Number of publications per year}\label{fig:paperperyear}
\end{figure}

The included studies were published between 2023 and 2025, reflecting the rapid growth of LLM-based approaches following the widespread availability of large-scale generative models. The temporal distribution of the reviewed studies shows a clear upward trend in research on LLM-based approaches for software engineering diagrams. As illustrated in Fig.~\ref{fig:paperperyear}, early work in 2023 was relatively limited, reflecting initial exploratory efforts in this area. Research activity increased substantially in 2024 and continued to grow in 2025, indicating sustained and accelerating interest.

\begin{figure}[!htbp]
    \centering
    \begin{subfigure}{0.48\textwidth}
        \centering        \includegraphics[width=\textwidth]{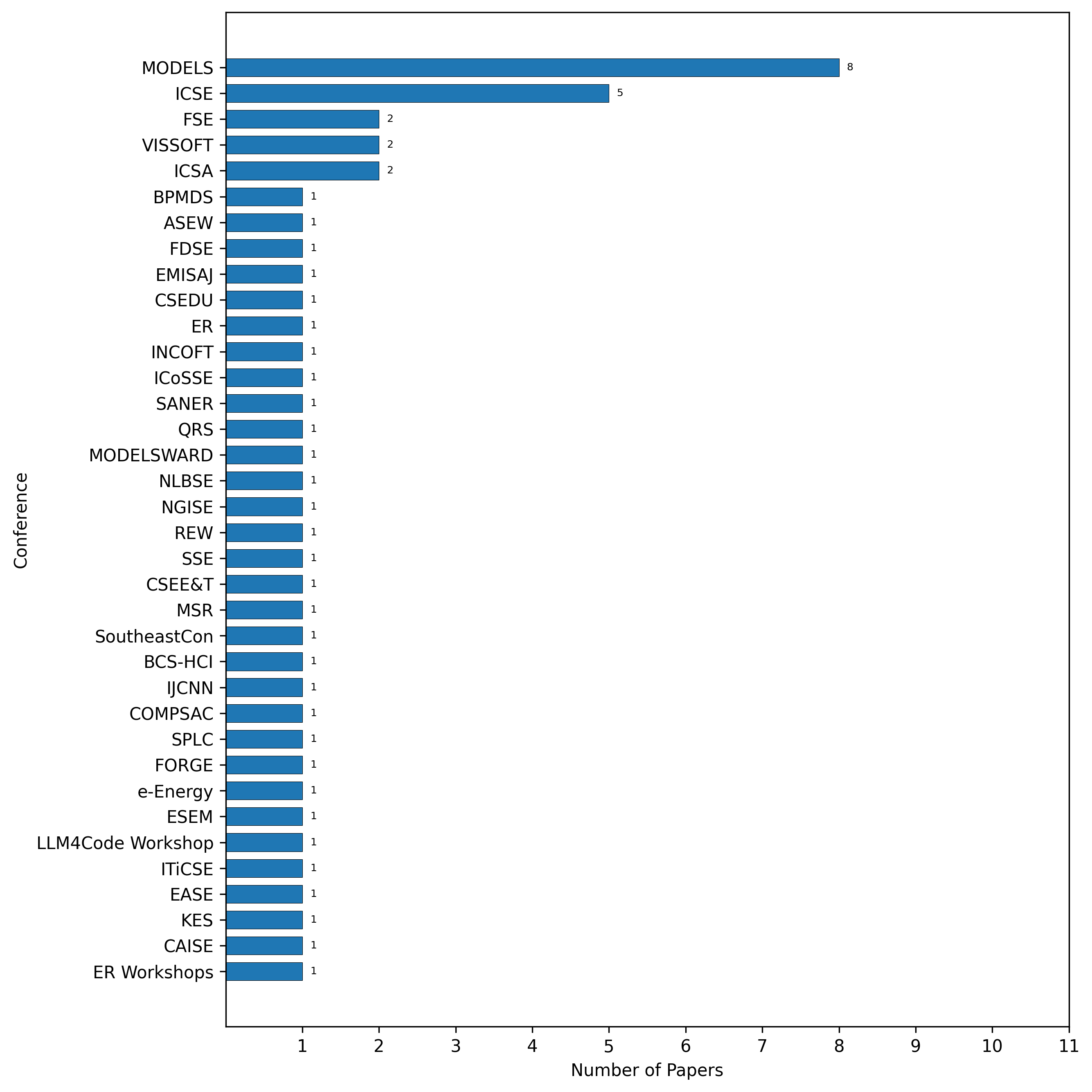}
        \caption{}
        \label{fig:a}
    \end{subfigure}
    \hfill
    \begin{subfigure}{0.48\textwidth}
        \centering        \includegraphics[width=\textwidth]{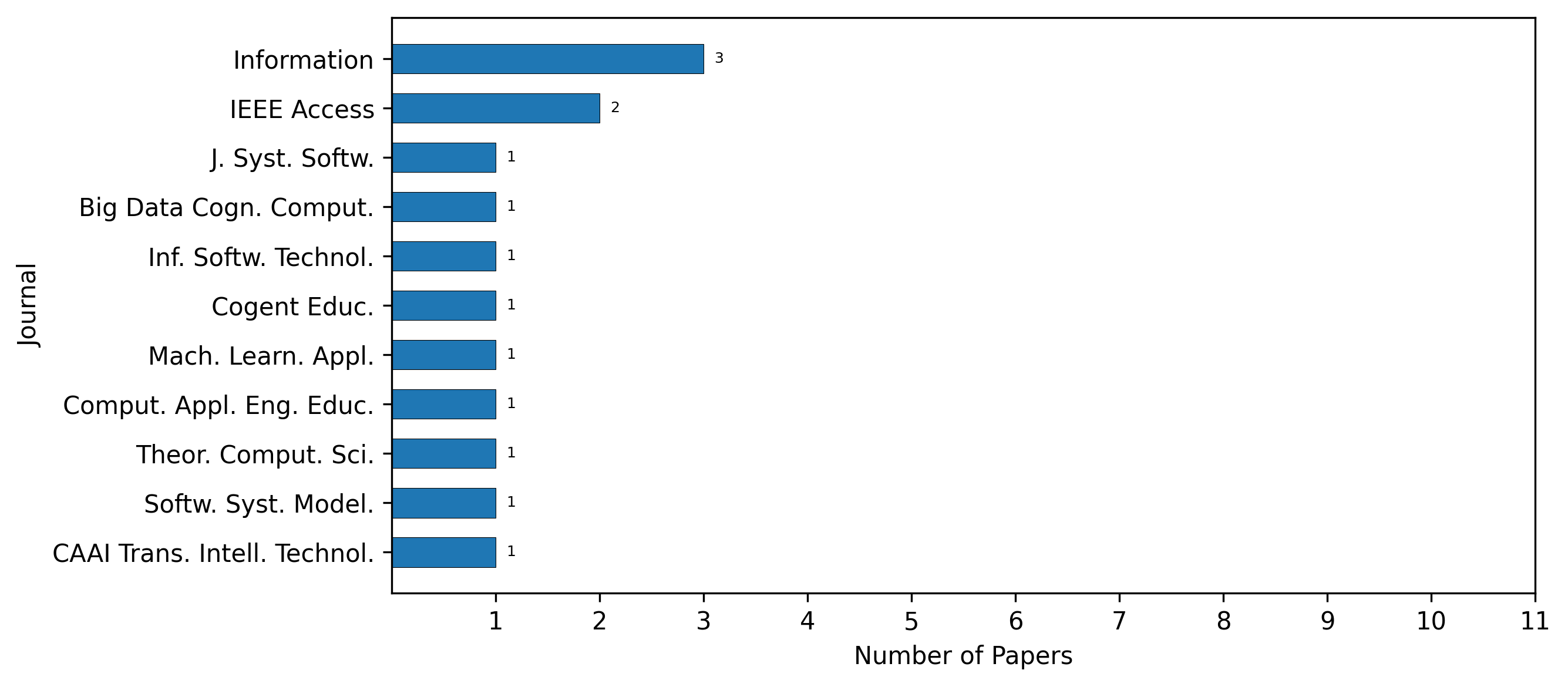}
        \caption{ }
        \label{fig:b}
    \end{subfigure}
    \caption{Distribution of the reviewed primary studies across (a) conference venues and (b) journal venues.}
    \label{fig:ab}
\end{figure}

Fig.~\ref{fig:a} and Fig.~\ref{fig:b} illustrate the distribution of primary studies across conference venues and journals, respectively. The results show that the majority of studies are published in well-established software engineering and model-driven engineering venues. Among conferences, International Conference on Model-Driven Engineering Languages and Systems (MODELS) is the most frequently represented venue, followed by IEEE/ACM International Conference on Software Engineering(ICSE), indicating strong alignment with the core MDE and SE research communities. A smaller number of studies are distributed across a wide range of specialized conferences and workshops, reflecting the interdisciplinary nature of the topic. With respect to journals, publications are more dispersed, with Information and IEEE Access hosting the largest number of studies, while other journals each contribute a limited number of papers.

\begin{figure}[!b]
\centering
\begin{adjustbox}{max width=\textwidth}
\begin{forest}
  for tree={
    font=\footnotesize,
    draw=RoyalBlue,
    align=left,
    edge={-latex},
    grow=east,
    reversed=true,
    parent anchor=east,
    child anchor=west,
    anchor=west,
    l sep=9mm,
    s sep=2mm,
    inner sep=2pt,
    rounded corners,
    edge path={
      \noexpand\path[\forestoption{edge}]
      (!u.parent anchor) -- +(6pt,0) |- (.child anchor);
    },
  }
[LLMs for SE Diagrams
  [\ref{subsec2:RQ1} Diagram Types
  [Software Modelling
    [UML Behavioral Diagrams
    [{Use Case Diagrams \\ 
\cite{reinhartz2025leveraging}, \cite{naimi2024automating}, \cite{tabassum2024using}, \cite{sultan2024ai}, \cite{o2024getting}, \cite{klimek2025re}, \cite{ramachandran2025transforming}, \cite{lu2025aug}, \cite{hassine2025evaluating}, \cite{wang2404llms}
 }]
    [{Sequence Diagrams \\ \cite{bates2025unified}, \cite{siala2024enhancing}, \cite{ouh2025evaluating}, \cite{jahan2024automated}, \cite{speth2024chatgpt}, \cite{lu2025aug}, \cite{xiao2025uml}, \cite{ferrari2024model}}]
    [{Activity Diagrams\\ \cite{zhao2024generating}, \cite{bates2025unified}, \cite{klimek2025re}, \cite{chaaben2023towards}, \cite{ben2024software}}]
    [{State Diagrams \\ \cite{moezkarimi2025harnessing}, \cite{ramachandran2025transforming}}]]
    [UML Structural Diagrams[{Class Diagram  \\\cite{camara2023assessment}, \cite{al2025student}, \cite{reinhartz2025leveraging}, \cite{ibanez2025can}, \cite{zhao2024generating}, \cite{yang2024multi}, \cite{wang2025devcoach}, \cite{siala2024enhancing}, \cite{ouh2025evaluating}, \\ \cite{de2024evaluating}, \cite{chaaben2023towards}, \cite{ben2024software}, \cite{ardimento2024enhancing}, \cite{ardimento2024rag}, \cite{antal2024toward}, \cite{ahmad2023towards}, \cite{acher2023generative}, \cite{abukhalaf2024pathocl}, \\ \cite{babaalla2025llm}, \cite{xue2024does}, \cite{li2024llm}, \cite{abukhalaf2023codex}, \cite{ramachandran2025transforming}, \cite{ardimento2024teaching}, \cite{shehata2024creating}, \cite{siala2025using}, \cite{wang2404llms},\\ \cite{lu2025aug}, \cite{al-ahmad_student-centric_2025}, \cite{ojha_towards_2025}, \cite{nair_fine-tuned_2025}, \cite{garaccione_evaluating_2025}, \cite{kop_evaluating_2025}, \cite{maass_application_2025}, \cite{bouali_toward_2025}, \\ \cite{fill_conceptual_2023} ,\cite{wang2025assessing} , \cite{chen2023automated},\cite{nguyen2025automated}
    }]
    [{Component Diagram  \\ \cite{fuchss2025lissa}, \cite{ahmad2023towards}, \cite{tagliaferro2025leveraging}}]
    [{Deployment Diagram \\  \cite{al2025student}, \cite{al-ahmad_student-centric_2025}}]
    ]
    ]
    [Data Modelling
    [{ER Diagrams \\  \cite{omar2023measurement}, \cite{rahmanian_challenges_2025}}]
    ]
  ]
  [\ref{subsec2:RQ2} Tasks Supported by LLMs
    [Diagram Construction from Natural Language
    [{Complete Diagram Generation \\ \cite{camara2023assessment}, \cite{reinhartz2025leveraging}, \cite{yang2024multi}, \cite{tabassum2024using}, \cite{sultan2024ai},\cite{kuchenbuch2025smart}, \cite{jahan2024automated}, \cite{de2024evaluating}, \cite{ahmad2023towards}, \cite{babaalla2025llm},\cite{ramachandran2025transforming} \\ \cite{omar2023measurement}, \cite{lu2025aug}, \cite{xiao2025uml}, \cite{tagliaferro2025leveraging}, \cite{ferrari2024model}, \cite{ojha_towards_2025}, \cite{nair_fine-tuned_2025}, \cite{garaccione_evaluating_2025} \\ \cite{maass_application_2025}, \cite{g_s_comparative_2025}, \cite{fill_conceptual_2023}, \cite{guizzardi_assessing_2025} , \cite{liu_ucd-llm_2026}, \cite{chen2023automated}, \cite{eisenreich2025leveraging}, \cite{nguyen2025automated}
    }]
    [{Partial diagram generation  \\ \cite{zhao2024generating}, \cite{klimek2025re}, \cite{li2024llm}}]
    ]
    [ Diagram Transformation 
    [{Diagram image to structured representation \\ \cite{bates2025unified}, \cite{hassine2025evaluating}}]
    [{Source code to UML models  (reverse engineering) \\ \cite{siala2024enhancing}, \cite{shehata2024creating}, \cite{siala2025using}}]
    [{Diagram/model to other formal artifacts \\ \cite{fuchss2025lissa}, \cite{antal2024toward}, \cite{moezkarimi2025harnessing}, \cite{abukhalaf2023codex}, \cite{gavric2024does}, \cite{acher2023generative}, \cite{abukhalaf2024pathocl}}]
    ]
    [{Automated or LLM-assisted assessment of \\ diagram correctness, quality, or grading \\  \cite{ibanez2025can}, \cite{rahmanian_challenges_2025}, \cite{bouali_toward_2025}, \cite{ramachandran_ai_2025}, \cite{wang2025assessing}}]
    [ Diagram Understanding and Human-Centric Assistance 
    [{Diagram Completion and Interactive Assistance \\ \cite{chaaben2023towards}, \cite{ben2024software}, \cite{ibanez2025can}, \cite{kop_evaluating_2025}}]
    [{Diagram Documentation and Explanation \\ \cite{naimi2024automating}, \cite{avignone_exploring_2025}}]
    [{Educational and Tutoring Support \\ \cite{al2025student}, \cite{ouh2025evaluating},\cite{o2024getting},  \cite{ardimento2024enhancing}, \cite{ardimento2024rag}, \cite{ardimento2024teaching}, \cite{yang2024deepocl}, \cite{wang2025devcoach}, \cite{xue2024does}, \cite{speth2024chatgpt},\cite{wang2404llms},\cite{al-ahmad_student-centric_2025}}]    
    ]
  ]    
  [\ref{subsec2RQ3} Technical Approaches]   
  [\ref{subsec2RQ4} Evaluation Metrics and Datasets]
] 
\end{forest}
\end{adjustbox}
\caption{Taxonomy of LLMs for Software Engineering Diagrams}
\label{fig:taxonomy_tree}
\end{figure}
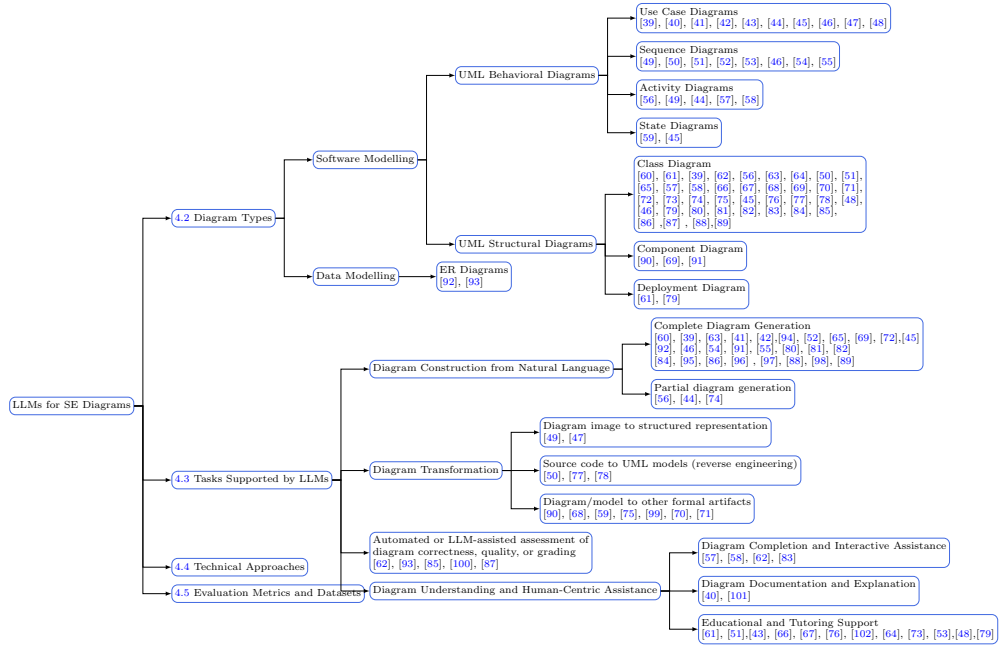

Fig.~\ref{fig:taxonomy_tree}  illustrates a multi-dimensional classification scheme of the reviewed studies, showing how papers are distributed across diagram types, tasks, techniques, evaluation approaches, and application domains.

\subsection{RQ1: Which types of UML and entity-relationship diagrams are addressed in LLM-based research? }\label{subsec2:RQ1}
This research question examines which types of software engineering diagrams have been addressed by existing LLM-based approaches. To ensure conceptual consistency and theoretical grounding, the classification of diagram types follows the classification defined in the The Unified modelling Language Reference Manual \cite{10.5555/993859}, which provides a canonical organization of UML diagrams based on their semantic focus and modelling intent. According to this classification, UML diagrams are grouped into behavioral and structural diagrams. Behavioral diagrams emphasize the dynamic aspects of a system, capturing interactions, workflows, and state-dependent behavior, whereas structural diagrams describe the static organization of system elements and their relationships. In addition to UML diagrams, the reviewed literature includes a subset of studies that focus on conceptual data models.

\begin{figure}[t]
\centering
\includegraphics[width=0.9\textwidth]{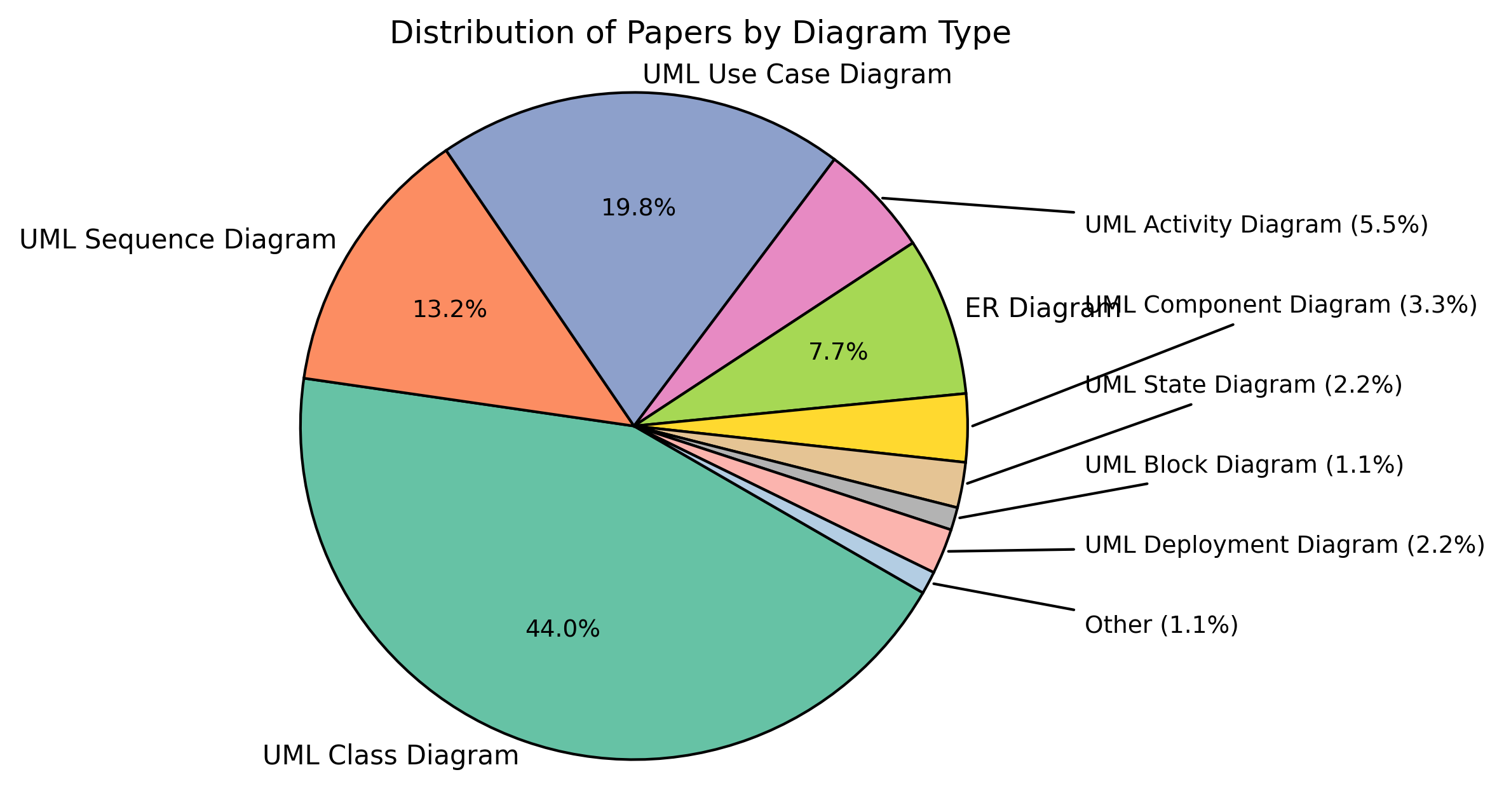}
\caption{Distribution of primary studies by software engineering diagram type}\label{fig:DisDType}
\end{figure}

Figure~\ref{fig:DisDType} illustrates the distribution of the reviewed studies across these diagram categories.

\subsubsection{RQ1.1: What types of behavioral diagrams are covered?}\label{RQ1.1}
 Use case diagrams receive the greatest attention among behavioral representations. A large subset of studies considers use case diagrams as primary modelling artifacts, focusing on actors, use cases, and their relationships as fundamental elements of system behavior \cite{reinhartz2025leveraging}, \cite{naimi2024automating}, \cite{tabassum2024using}, \cite{sultan2024ai}, \cite{o2024getting}, \cite{klimek2025re}, \cite{ramachandran2025transforming}, \cite{lu2025aug}, \cite{hassine2025evaluating}, \cite{wang2404llms}. Their prevalence highlights the role of use case diagrams as high-level abstractions for representing user–system interactions and defining system boundaries.

Sequence diagrams represent another prominent behavioral diagram type in the reviewed corpus. Several studies address interaction-oriented behavioral representations involving lifelines, messages, and their ordering \cite{bates2025unified}, \cite{siala2024enhancing}, \cite{ouh2025evaluating}, \cite{jahan2024automated}, \cite{speth2024chatgpt}, \cite{lu2025aug}, \cite{xiao2025uml}, \cite{ferrari2024model}. 

Activity diagrams are addressed less frequently and are often used as intermediate representations within broader modelling pipelines rather than as primary outputs. In these studies, LLMs are employed to decompose textual narratives into action flows and control structures \cite{zhao2024generating}, \cite{bates2025unified}, \cite{klimek2025re}, \cite{chaaben2023towards}, \cite{ben2024software}. 

State machine diagrams appear only in a small number of studies \cite{moezkarimi2025harnessing}, \cite{ramachandran2025transforming}. These works consider state-based behavioral representations centered on states, transitions, and triggering events. In contrast to interaction-oriented behavioral diagrams, state machine diagrams remain underrepresented in the reviewed literature.

\subsubsection{ RQ1.2: What types of structural diagrams (UML and ER) are covered?}\label{RQ1.2}
Several families of structural diagrams exist, but their representation across the corpus is highly uneven. A clear majority of studies focus on class-based modelling. By contrast, other structural diagram types, including component, deployment, and ER diagrams, are much less frequently represented.

Class diagrams are the dominant structural diagram type in the corpus. A substantial number of works consider class diagrams as primary modelling representations, focusing on classes, attributes, operations, and relationships such as associations, generalizations, and compositions. (see Fig.~\ref{fig:taxonomy_tree} for a complete overview of the corresponding studies ).

Component diagrams are addressed in a smaller subset of studies, indicating more limited coverage of structural representations at the architectural component level \cite{fuchss2025lissa}, \cite{ahmad2023towards}, \cite{tagliaferro2025leveraging}. 

Deployment diagrams are considered in only a small subset of the reviewed literature and are therefore underrepresented relative to class-based structural modelling \cite{al2025student}, \cite{al-ahmad_student-centric_2025}.

ER diagrams are addressed in a limited subset of the reviewed literature. ER models emphasize entities, attributes, relationships, and cardinality constraints, and they are typically evaluated with respect to schema correctness and data consistency rather than software design properties. The identified studies apply LLM-based approaches to tasks involving ER diagram generation and assessment. In particular, existing work explores the derivation of ER models from natural-language specifications as well as the evaluation of ER diagrams produced by human modelers \cite{omar2023measurement}, \cite{rahmanian_challenges_2025}. These approaches leverage LLMs’ ability to interpret domain terminology and relational descriptions, enabling the identification of core entities and candidate relationships even when input descriptions are informal or incomplete.

\subsection{RQ2: What diagram-centric tasks are supported by LLM-based approaches?
}\label{subsec2:RQ2}
To answer RQ2, we examine the types of diagram-centric tasks supported by LLM-based approaches in software modelling contexts. Specifically, we analyse the modelling activities that LLMs are employed to perform, independently of the particular diagram type or technical implementation strategy. RQ2 is structured into two subquestions:

\indent \begin{itemize}
    \item \textbf{RQ2.1:} What categories of diagram-centric tasks are addressed by existing studies? 
    \item \textbf{RQ2.2:} What input and output types are associated with these tasks?
\end{itemize}

\subsubsection{RQ2.1: What categories of diagram-centric tasks are addressed by existing studies?}\label{RQ2.1}
To address RQ2.1, we analyse the categories of diagram-centric tasks supported by LLM-based approaches in the reviewed studies. Rather than focusing on specific diagram types or implementation strategies, this analysis examines the primary functional objectives assigned to LLMs within modelling activities. Based on a synthesis of the corpus, we identify four main task categories: (1) diagram construction, (2) diagram transformation, (3) diagram quality assurance, and (4) diagram understanding and assistance. The following subsections describe each task category in detail.

\begin{table*}[t]
\centering
\small
\caption{Task Categories in LLM-Based Diagram-Centric Studies}
\label{tab:rq2_Tasks}
\setlength{\tabcolsep}{4pt}
\begin{tabular}{ p{0.33\linewidth} p{0.38\linewidth} p{0.24\linewidth}}
\toprule
\textbf{Type} & \textbf{Description} & \textbf{Papers} \\
\midrule

\multicolumn{3}{l}{\textit{Diagram Construction from Natural Language}} \\
\midrule
Complete Diagram Generation 
& Direct generation of complete diagrams from natural-language requirements.
& \cite{camara2023assessment}, \cite{reinhartz2025leveraging}, \cite{yang2024multi}, \cite{tabassum2024using}, \cite{sultan2024ai},\cite{kuchenbuch2025smart}, \cite{jahan2024automated}, \cite{de2024evaluating}, \cite{ahmad2023towards}, \cite{babaalla2025llm},\cite{ramachandran2025transforming}, \cite{omar2023measurement}, \cite{lu2025aug}, \cite{xiao2025uml}, \cite{tagliaferro2025leveraging}, \cite{ferrari2024model}, \cite{ojha_towards_2025}, \cite{nair_fine-tuned_2025}, \cite{garaccione_evaluating_2025}, \cite{maass_application_2025}, \cite{g_s_comparative_2025}, \cite{fill_conceptual_2023}, \cite{guizzardi_assessing_2025} , \cite{liu_ucd-llm_2026}, \cite{chen2023automated}, \cite{eisenreich2025leveraging}, \cite{nguyen2025automated}\\

\addlinespace

Partial Diagram Generation 
& Generation of partial diagram elements requiring refinement.
& \cite{zhao2024generating}, \cite{klimek2025re}, \cite{li2024llm}  \\

\midrule
\multicolumn{3}{l}{\textit{Diagram Transformation}} \\
\midrule
Diagram image to structured representation 
& Extraction of structured model elements from diagram images.
& \cite{bates2025unified}, \cite{hassine2025evaluating} \\
\addlinespace
Source code to UML models 
& Reverse engineering from source code to diagram artifacts.
& \cite{siala2024enhancing}, \cite{shehata2024creating}, \cite{siala2025using} \\
\addlinespace
Diagram/model to other formal artifacts 
& Conversion between modelling artifacts or derivation of formal representations.
& \cite{fuchss2025lissa}, \cite{antal2024toward}, \cite{moezkarimi2025harnessing}, \cite{abukhalaf2023codex}, \cite{gavric2024does}, \cite{acher2023generative}, \cite{abukhalaf2024pathocl} \\

\midrule
\multicolumn{3}{l}{\textit{Diagram Evaluation}} \\
\midrule
Diagram Evaluation 
& Automated or LLM-assisted assessment of diagram correctness, quality, or grading.
&  \cite{ibanez2025can}, \cite{rahmanian_challenges_2025}, \cite{bouali_toward_2025}, \cite{ramachandran_ai_2025}, \cite{wang2025assessing}\\

\midrule
\multicolumn{3}{l}{\textit{Diagram Understanding and Human-Centric Assistance}} \\
\midrule
Diagram Completion and Interactive Assistance 
& Interactive support, refinement, or completion of modelling artifacts.
& \cite{chaaben2023towards}, \cite{ben2024software}, \cite{ibanez2025can}, \cite{kop_evaluating_2025}\\
\addlinespace
Diagram Documentation and Explanation 
& Natural-language documentation or explanation of diagram artifacts.
& \cite{naimi2024automating}, \cite{avignone_exploring_2025} \\
\addlinespace
Educational and Tutoring Support 
& Grading, instructional feedback, and learning-oriented modelling assistance.
& \cite{al2025student} \cite{ouh2025evaluating},\cite{o2024getting},  \cite{ardimento2024enhancing}, \cite{ardimento2024rag}, \cite{ardimento2024teaching}, \cite{yang2024deepocl}, \cite{wang2025devcoach}, \cite{xue2024does},\cite{speth2024chatgpt},\cite{wang2404llms},\cite{al-ahmad_student-centric_2025} \\

\bottomrule
\end{tabular}
\end{table*}

\textbf{\textit{1) Diagram Construction:}} In this review, diagram construction from natural language refers to approaches in which a large language model receives a textual description, such as requirements statements, user stories, or domain narratives, as input and produces diagrammatic artifacts as output. The resulting artifacts are typically expressed in machine readable diagram representations, such as PlantUML or Mermaid specifications, serialized UML or ER model encodings, or structured textual representations that can be rendered into diagrams. 
    
\textbf{Complete diagram generation} refers to approaches in which a large language model produces a complete and self-contained diagram representation directly from a natural-language description. Typical outputs include full class diagrams, sequence diagrams, use case diagrams, and entity–relationship diagrams encoded in textual diagram notations that can be rendered into standard modelling views. In this category, the diagram constitutes the primary end product of the task, even when subsequent validation or refinement steps are incorporated after generation. As summarized in Table \ref{tab:rq2_Tasks}, a substantial number of studies investigate this form of text-driven diagram construction \cite{camara2023assessment}, \cite{reinhartz2025leveraging}, \cite{de2024evaluating}, \cite{omar2023measurement}, \cite{xiao2025uml}, \cite{tabassum2024using}, \cite{ahmad2023towards}, \cite{ferrari2024model}.

\textbf{Partial diagram generation} refers to approaches in which a large language model produces diagram elements or intermediate modelling artifacts rather than a complete diagram. The outputs typically include identified classes or entities, extracted relationships, domain concepts, or other structured representations derived from natural language descriptions. In these studies, the generated artifacts contribute to diagram construction but do not constitute a fully specified model. Studies \cite{zhao2024generating}, \cite{klimek2025re}, and \cite{li2024llm} follow this approach, where the model supports early or intermediate modelling activities by externalizing key elements.

\begin{figure}[t]
\centering
\includegraphics[width=0.7\textwidth]{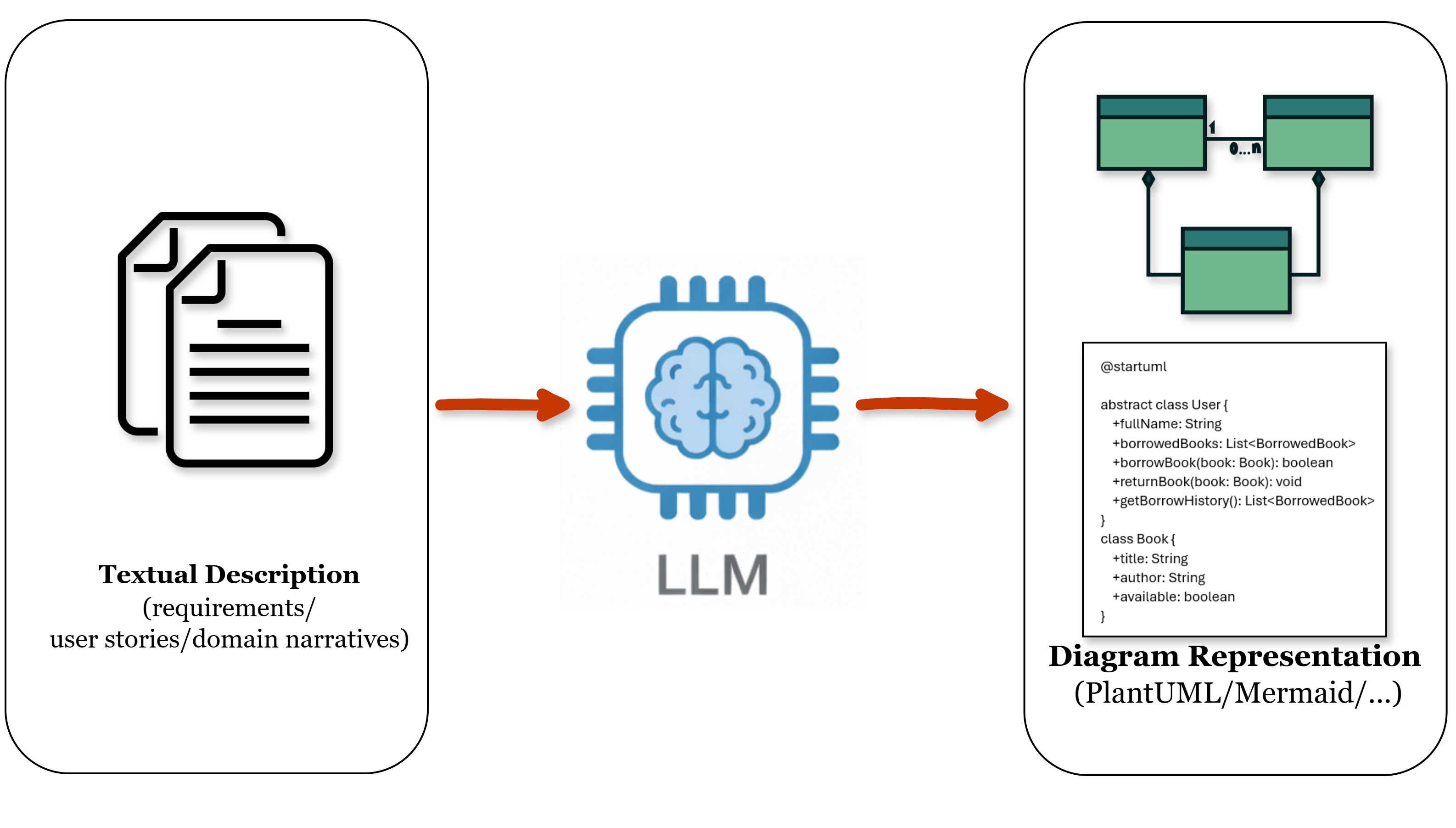}
\caption{Diagram construction tasks supported by large language models.}\label{fig:Task1}
\end{figure}

\textbf{\textit{2) Diagram transformation:}}
Diagram transformation constitutes a major task category in which LLMs are used to translate existing software artifacts into alternative representations. In contrast to diagram construction, which typically involves generating models from textual descriptions, transformation tasks operate on pre-existing diagrams or related artifacts as input. The objective of these tasks is to produce alternative diagrammatic, textual, or code-level representations derived from an existing model.

\begin{figure}[t]
\centering
\includegraphics[width=0.7\textwidth]{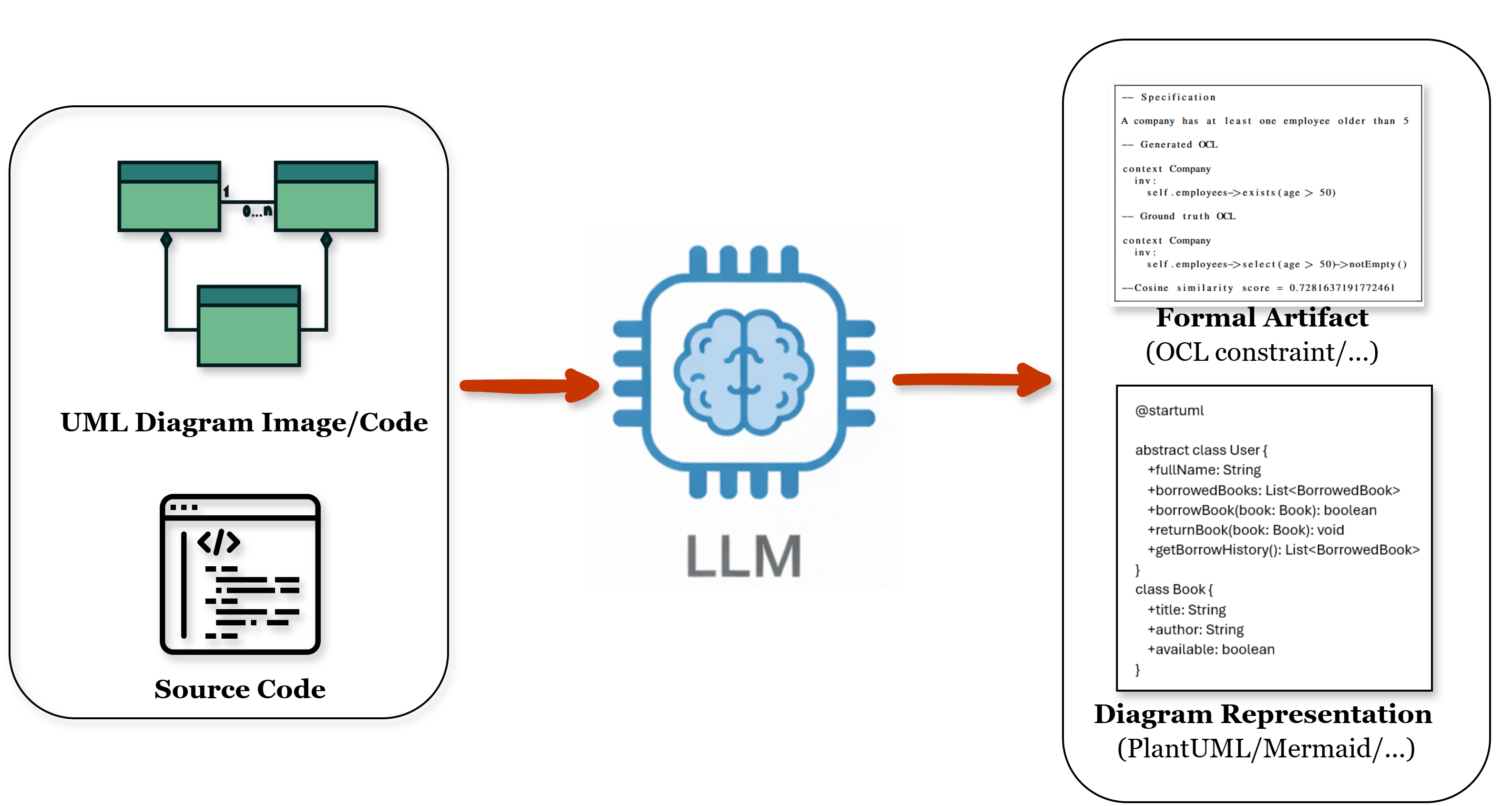}
\caption{Diagram transformation tasks supported by large language models}\label{fig:Task2}
\end{figure}

\textbf{Diagram image to structured representation} refers to transformation tasks in which a large language model processes diagram images as input and produces structured, machine readable representations as output. In these approaches, the input typically consists of visual depictions of UML or related diagrams, such as screenshots or scanned images, while the output comprises structured diagram elements or serialized model representations that can be further analysed or rendered. In the reviewed studies, this transformation is typically supported by multimodal large language models that combine visual perception with language based reasoning to interpret diagram images and recover structured representations \cite{bates2025unified}, \cite{hassine2025evaluating}. In this setting, the diagram image serves as the primary artifact, and the structured model representation constitutes the transformation output.

\textbf{Source code to UML model transformation} is addressed in a limited number of studies that frame the task as reverse engineering from implementation artifacts to design-level models. In this setting, object-oriented source code serves as the primary input, and UML representations constitute the output. Study \cite{siala2024enhancing} applies large language models to object-oriented source code to derive UML class diagram elements, including classes, attributes, and associations, with an emphasis on semantic abstraction rather than direct syntactic extraction. Study \cite{shehata2024creating} focuses on Java source code and employs LLMs to recover UML class diagrams intended to reflect architectural structure and design intent. In \cite{siala2025using}, LLMs are combined with manually defined abstraction rules to transform object-oriented source code, evaluated primarily on Java, into UML models.

\textbf{Diagram and model to formal artifact} transformation refers to approaches in which a large language model takes diagrammatic or structured model representations as input and produces other formal artifacts as output. In this category, diagrams and models are reinterpreted into representations intended for analysis, verification, or execution rather than being treated as final modelling products.

The reviewed studies illustrate a range of transformation targets. These include the derivation of formal constraints from models \cite{antal2024toward}, the generation of executable or analyzable specifications from diagrammatic representations \cite{moezkarimi2025harnessing}, and the translation of models into artifacts supporting verification or consistency checking \cite{abukhalaf2023codex}. Other works apply LLMs to domain-specific or architectural models to produce structured formal representations governed by predefined modelling viewpoints \cite{gavric2024does}, or to transform models into analysis-oriented artifacts that support traceability and semantic reasoning \cite{acher2023generative}, \cite{abukhalaf2024pathocl}.

\textbf{\textit{3) Diagram evaluation:}} Diagram evaluation refers to approaches in which a large language model takes a diagram or structured model representation as input and produces evaluative outputs rather than modifying the diagram itself. In these tasks, the LLM is employed to assess aspects such as quality, correctness, or completeness, with outputs ranging from numerical grades or categorical judgments to more detailed evaluative explanations.

As summarized in Table \ref{tab:rq2_Tasks},  the reviewed studies apply LLMs to the evaluation of UML or conceptual diagrams in both educational and non-educational contexts. Several works focus on grading or scoring diagrams against reference solutions or predefined criteria, producing assessment outcomes without accompanying diagnostic feedback \cite{ibanez2025can}, \cite{rahmanian_challenges_2025}, \cite{bouali_toward_2025}. In contrast, study \cite{ramachandran_ai_2025} positions the LLM as a reviewer that supplements evaluative judgments with qualitative explanations intended to justify assessment decisions.

\begin{figure}[t]
\centering
\includegraphics[width=0.8\textwidth]{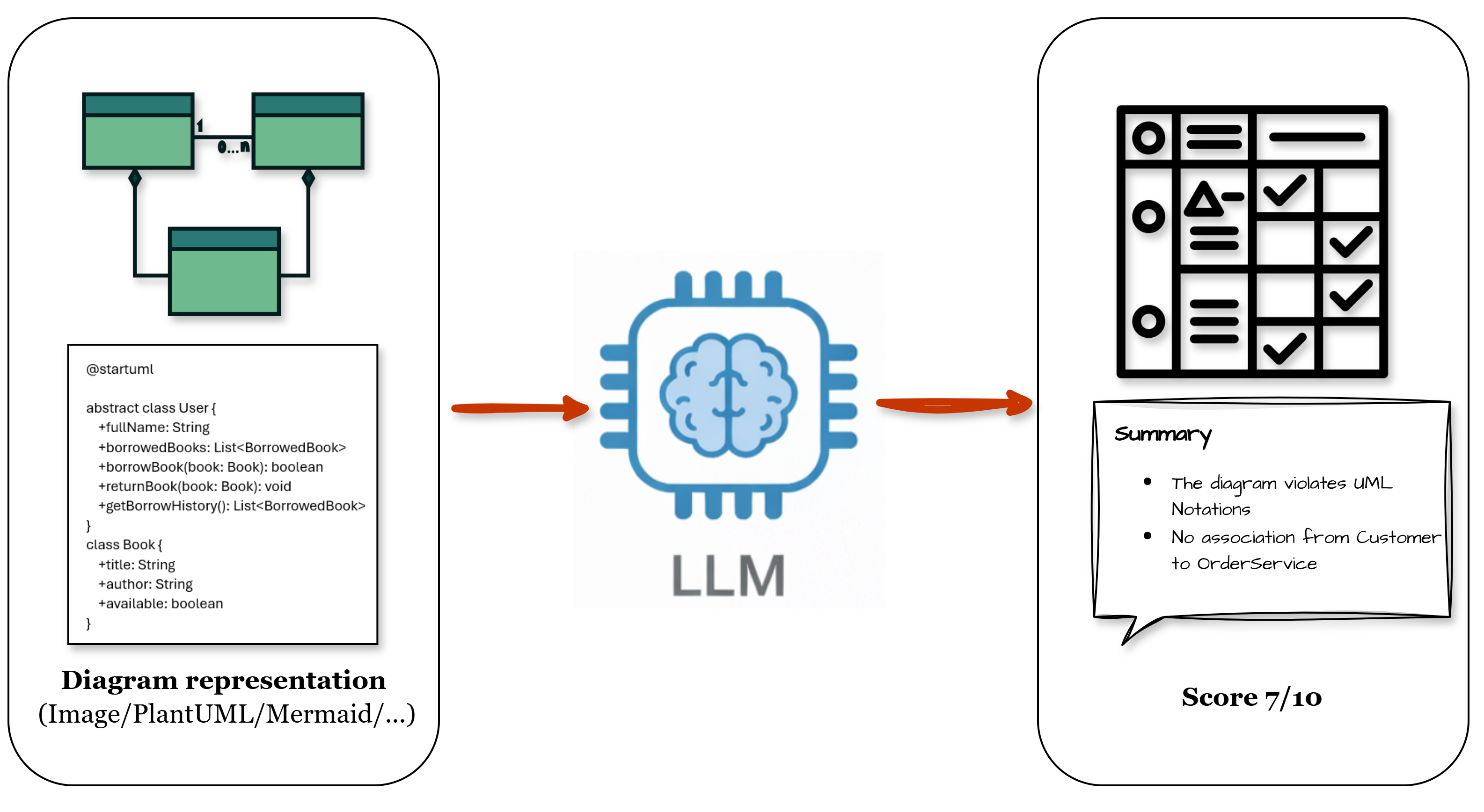}
\caption{Diagram evaluation tasks supported by large language models.}\label{fig:Task3}
\end{figure}

\textbf{\textit{4) Diagram understanding and assistance:}} This category encompasses tasks in which large language models operate on existing diagram artifacts to support user comprehension, interaction, or learning activities. Diagrams constitute the primary input, and the outputs provide explanatory, assistive, or instructional support.
As summarized in Table \ref{tab:rq2_Tasks}, the reviewed studies in this category can be grouped into three subtypes: diagram completion and interactive assistance, diagram documentation and explanation, and educational or tutoring support. These tasks position the LLM as an interactive assistant operating on diagram artifacts to support user-oriented activities.

\textbf{Diagram completion and interactive assistance} refer to approaches in which a large language model operates on an existing diagram to support user interaction or understanding, rather than generating a new diagram from textual input. In this category, the input typically consists of an incomplete or partially specified diagram, or a complete diagram queried for specific information. The output includes completed diagram elements, extracted components or relationships, or structured responses that assist users in navigating and interpreting the diagram.

Some studies focus on diagram completion, where LLMs infer and populate missing elements or relationships in partially specified diagrams based on contextual or structural cues \cite{chaaben2023towards}, \cite{ben2024software}. Other works emphasize interactive access to diagram content, applying LLMs to extract entities, relationships, or diagram fragments in response to user queries \cite{ibanez2025can}, \cite{kop_evaluating_2025}.

\textbf{Diagram documentation and explanation} refer to approaches in which a large language model generates natural-language descriptions of existing diagrams to support comprehension, communication, or documentation. In these tasks, the input consists of a diagram or structured diagram representation, and the output is a textual explanation summarizing its elements, relationships, or overall intent.

The reviewed studies apply LLMs to transform UML or conceptual diagrams into descriptive narratives that capture structural and semantic aspects of the model. Some works focus on producing documentation-oriented descriptions, such as concise summaries or structured explanations intended to accompany design artifacts and improve their accessibility for stakeholders \cite{naimi2024automating}. Other studies emphasize interpretive explanations, where LLMs generate more detailed natural-language accounts aimed at clarifying diagram semantics or supporting understanding by non-expert readers \cite{avignone_exploring_2025}.

\textbf{Educational and tutoring support} encompass approaches in which a large language model operates on diagrammatic artifacts to assist learning, instruction, or skill development in modelling tasks. In this category, diagrams, often produced by students or learners, serve as the primary input. The outputs include guidance, explanations, hints, or instructional responses intended to support comprehension and learning progression.

Several studies investigate the use of LLMs as intelligent tutors that provide step-by-step guidance or respond to questions related to UML or conceptual diagrams in educational settings \cite{ouh2025evaluating}, \cite{ardimento2024enhancing}, \cite{ardimento2024rag}. Other works focus on learning-oriented assistance, where LLMs generate explanations or contextual guidance tailored to learner needs \cite{yang2024deepocl}, \cite{wang2025devcoach}, \cite{speth2024chatgpt}. Additional studies explore the use of LLMs to support diagram-based learning activities, such as assisting students in interpreting or reasoning about diagrams within software engineering coursework \cite{ardimento2024teaching}.

 \subsubsection{RQ2.2: What input and output types are associated with these tasks?}\label{RQ2.2}
The reviewed studies employ a range of input modalities, primarily centered on textual and structural artifacts. As shown in Table~\ref{tab:rq2_input}, natural-language requirements and domain descriptions constitute the most common input type. These inputs typically include user stories, functional specifications, or informal system descriptions used to generate or transform modelling artifacts. Source code is also used in a subset of studies, particularly in reverse engineering. Image-based diagram inputs appear mainly in grading and extraction scenarios, while serialized model files and multimodal combinations (e.g., text combined with rubric or diagram artifacts) are less frequent but present.

The outputs produced by LLM-based approaches are similarly diverse. Table~\ref{tab:rq2_output} summarizes the main output artifact types. The most common output form is structured textual model representation, including machine-readable encodings of diagrams or models. Rendered diagram images are reported in studies that prioritize visual output. Other outputs include formal constraints (e.g., OCL), automated grades or scores, textual explanations or feedback, and query expressions.

\begin{table*}[t] \centering \small \caption{Input types used in LLM-based diagram-centric tasks} \label{tab:rq2_input} \begin{tabular}{p{0.30\linewidth} p{0.35\linewidth} p{0.25\linewidth}} \toprule \textbf{Input Type} & \textbf{Description} & \textbf{Papers} \\ \midrule \textbf{Natural-language requirements / domain descriptions} & Free-form textual requirements, system scenarios, user stories, functional specifications, or domain descriptions provided as input for diagram generation or transformation. &  \cite{camara2023assessment}, \cite{al2025student}, \cite{reinhartz2025leveraging}, \cite{yang2024deepocl},  \cite{zhao2024generating}, \cite{yang2024multi}, \cite{wang2025devcoach}, \cite{tabassum2024using}, \cite{sultan2024ai}, \cite{klimek2025re}, \cite{jahan2024automated}, \cite{de2024evaluating}, \cite{ben2024software}, \cite{ahmad2023towards}, \cite{abukhalaf2024pathocl}, \cite{babaalla2025llm}, \cite{li2024llm}, \cite{speth2024chatgpt}, \cite{abukhalaf2023codex}, \cite{ramachandran2025transforming}, \cite{omar2023measurement}, \cite{xiao2025uml}, \cite{wang2404llms}, \cite{tagliaferro2025leveraging}, \cite{ferrari2024model}, \cite{al-ahmad_student-centric_2025}, \cite{ojha_towards_2025}, \cite{nair_fine-tuned_2025}, \cite{garaccione_evaluating_2025}, \cite{kop_evaluating_2025}, \cite{maass_application_2025}, \cite{g_s_comparative_2025}, \cite{fill_conceptual_2023}, \cite{guizzardi_assessing_2025}, \cite{liu_ucd-llm_2026}, \cite{chen2023automated}, \cite{eisenreich2025leveraging}, \cite{nguyen2025automated} \\ \addlinespace \textbf{Source code} & Source code artifacts (e.g., Java or Python files) & \cite{siala2024enhancing}, \cite{fuchss2025lissa}, \cite{shehata2024creating}, \cite{siala2025using} \\ \addlinespace \textbf{Diagram image} & Image-based diagram inputs (e.g., PNG/SVG of UML or ER diagrams) & \cite{ibanez2025can}, \cite{bates2025unified}, \cite{ouh2025evaluating}, \cite{hassine2025evaluating}, \cite{rahmanian_challenges_2025}, \cite{bouali_toward_2025}, \cite{wang2025assessing} \\ \addlinespace \textbf{Textual model representation} & Machine-readable structural representations of models, including diagram encodings (e.g., PlantUML), model files (e.g., XMI), and structured element lists (e.g., JSON-based entities and relationships). & \cite{o2024getting}, \cite{chaaben2023towards}, \cite{moezkarimi2025harnessing}, \cite{ardimento2024teaching}, \cite{lu2025aug},\cite{naimi2024automating}, \cite{ardimento2024enhancing}, \cite{ardimento2024rag}, \cite{antal2024toward}, \cite{avignone_exploring_2025} \\ \addlinespace \textbf{Mixed or multimodal inputs} & Combined inputs such as text + diagram, code + model artifacts. & \cite{kuchenbuch2025smart}, \cite{xue2024does}, \cite{ramachandran_ai_2025} \\ \bottomrule \end{tabular} \end{table*}

\begin{table*}[t]
\centering
\small
\caption{Output types produced in LLM-based diagram-centric tasks}
\label{tab:rq2_output}
\begin{tabular}{p{0.30\linewidth} p{0.35\linewidth} p{0.25\linewidth}}
\toprule
\textbf{Output Type} & \textbf{Description} & \textbf{Papers} \\
\midrule

Textual model representation
& Machine-readable structural representations of models, including diagram encodings (e.g., PlantUML), model files (e.g., XMI), and structured element lists (e.g., JSON-based entities and relationships). 
& \cite{camara2023assessment}, \cite{al2025student}, \cite{reinhartz2025leveraging}, \cite{bates2025unified},  \cite{ouh2025evaluating}, \cite{jahan2024automated}, \cite{de2024evaluating}, \cite{ahmad2023towards}, \cite{acher2023generative}, \cite{ramachandran2025transforming}, \cite{shehata2024creating}, \cite{xiao2025uml}, \cite{wang2404llms}, \cite{tagliaferro2025leveraging}, \cite{ferrari2024model}, \cite{al-ahmad_student-centric_2025}, \cite{ojha_towards_2025}, \cite{nair_fine-tuned_2025}, \cite{garaccione_evaluating_2025}, \cite{kop_evaluating_2025}, \cite{maass_application_2025}, \cite{g_s_comparative_2025}, \cite{fill_conceptual_2023}, \cite{guizzardi_assessing_2025}, \cite{liu_ucd-llm_2026}, \cite{eisenreich2025leveraging}, \cite{nguyen2025automated}, \cite{tabassum2024using}, \cite{sultan2024ai}, \cite{o2024getting}, \cite{kuchenbuch2025smart}, \cite{klimek2025re}, \cite{fuchss2025lissa}, \cite{chaaben2023towards}, \cite{ben2024software}, \cite{ardimento2024enhancing}, \cite{ardimento2024rag}, \cite{antal2024toward}, \cite{moezkarimi2025harnessing}, \cite{babaalla2025llm}, \cite{ardimento2024teaching}, \cite{lu2025aug}, \cite{hassine2025evaluating}, \cite{rahmanian_challenges_2025}, \cite{bouali_toward_2025}, \cite{chen2023automated} \\

\addlinespace
Diagram image
& Visual diagram output (e.g., rendered UML/ER diagram image such as PNG/SVG).
& \cite{gavric2024does}, \cite{ibanez2025can}, \cite{speth2024chatgpt}, \cite{abukhalaf2023codex}, \cite{omar2023measurement}, \cite{siala2025using}, \cite{wang2025assessing} \\

\addlinespace
Formal constraints (OCL)
& Generated formal constraints, primarily in OCL.
& \cite{yang2024deepocl}, \cite{zhao2024generating}, \cite{siala2024enhancing}, \cite{abukhalaf2024pathocl}, \cite{avignone_exploring_2025} \\

\addlinespace
Grades / scores
& Numerical assessment outputs (e.g., rubric-based grades/scores)
& \cite{naimi2024automating}, \cite{xue2024does}, \cite{li2024llm}, \cite{ramachandran2025transforming}, \cite{rahmanian_challenges_2025}, \cite{maass_application_2025}, \cite{bouali_toward_2025}, \cite{ramachandran_ai_2025}, \cite{wang2025assessing} \\

\addlinespace
Textual feedback / explanation
& Natural-language documentation, explanations, or feedback generated
& \cite{yang2024multi}, \cite{wang2025devcoach},
\cite{babaalla2025llm}, \cite{abukhalaf2023codex}, \cite{shehata2024creating}, \cite{hassine2025evaluating}, \cite{siala2025using}, \cite{xiao2025uml}, \cite{avignone_exploring_2025}, \cite{fill_conceptual_2023}, \cite{eisenreich2025leveraging} \\

\addlinespace
Diagram artifact
& Diagram output reported, but the form (rendered image vs.\ serialized encoding) is not specified in the study.
& \cite{shehata2024creating}, \cite{al-ahmad_student-centric_2025} \\
\bottomrule
\end{tabular}
\end{table*}

\subsection{RQ3: What LLM techniques and technical configurations are employed in diagram-centric modelling?}\label{subsec2RQ3}
To answer RQ3, we examine the technical approaches through which large language models are configured and integrated in diagram-centric modelling studies. While RQ2 focuses on the functional tasks supported by LLMs, RQ3 shifts attention to the underlying techniques, model configurations, and system-level design choices that enable these tasks.

RQ3 is structured into three subquestions:

\indent \begin{itemize}
    \item \textbf{RQ3.1:} What LLM models and configuration strategies are used in diagram-centric modelling studies?
    \item \textbf{RQ3.2:} 
    What prompting techniques are used to guide LLM outputs?
    \item \textbf{RQ3.3:}  What architectural and workflow patterns are used to integrate LLMs?
\end{itemize}

 \subsubsection{RQ3.1: What LLM models and configuration strategies are used in diagram-centric modelling studies?}\label{RQ3.1}
To address this research question, we examine the distribution of model families and configuration strategies reported in the reviewed studies.

\begin{figure}
\centering
\includegraphics[width=1\textwidth]{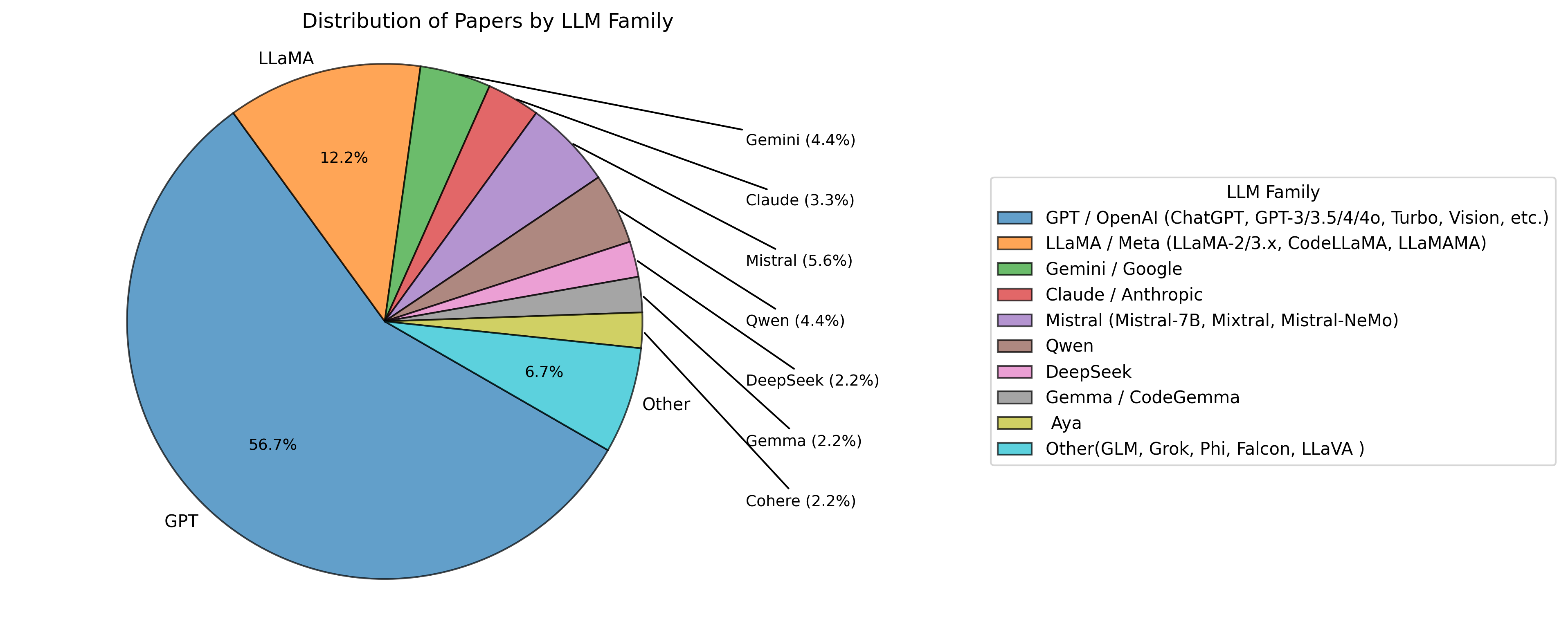}
\caption{LLM Models}\label{fig:Models}
\end{figure}

\textbf{LLM Model Families:} The corpus shows a strong concentration around a limited number of foundation model providers, with a clear dominance of GPT-based models (see Figure \ref{fig:Models},  and Table X for full references).

\textbf{Fine-Tuning and Model Adaptation:} A small subset of the reviewed studies explores fine-tuning large language models to adapt them to software-modelling–related tasks. Unlike prompt-based or retrieval-augmented approaches, fine-tuning operates by updating model parameters using task-specific data, thereby embedding modelling knowledge directly into the model. Due to the computational cost and data requirements involved, all identified studies adopt parameter-efficient fine-tuning (PEFT) strategies rather than full model retraining.

Across the reviewed literature, Low-Rank Adaptation (LoRA) and its variants constitute the dominant fine-tuning mechanism. In these approaches, lightweight trainable rank-decomposition matrices are injected into selected transformer layers while the base model parameters remain frozen, enabling efficient specialization with limited resources. For example, \cite{bates2025unified} applies LoRA to adapt a pretrained LLM for UML diagram generation in an educational context, focusing on internalizing diagrammatic structures through supervised fine-tuning. Similarly, \cite{nair_fine-tuned_2025} employs QLoRA, a quantized variant of LoRA, to fine-tune an LLM for extracting UML class diagrams from source code, combining parameter-efficient adaptation with reduced memory overhead.

Fine-tuning is also used in tasks adjacent to diagram generation, where models are adapted to produce model-driven artifacts or queries. In [25], LoRA is used to fine-tune open-source LLMs for generating expressions in a model query language, leveraging synthetic and curated training data to teach the model formal query constructs. Although the task does not involve direct diagram synthesis, the approach demonstrates how parameter-efficient fine-tuning can be applied to model-centric languages and artifacts within the broader model-driven engineering landscape.

In some studies, fine-tuning is embedded within larger analysis pipelines rather than used as a standalone solution. For instance, \cite{siala2025using} applies supervised fine-tuning with QLoRA to specialize an LLM for extracting UML class diagrams from Java programs, while relying on additional abstraction rules and post-processing steps to structure the final models. In this setting, fine-tuning serves to align the LLM with domain-specific modelling patterns, while deterministic components handle rule enforcement and representation constraints.

  \subsubsection{RQ3.2:     What prompting techniques are used to guide LLM outputs?}\label{RQ3.2}
  
To address RQ3.2, we analyse the prompting and interaction strategies reported in the reviewed studies for guiding LLM behavior in diagram-centric modelling tasks. This section categorizes three prompting techniques identified in the corpus and references the corresponding studies.

\textbf{1) Zero-shot prompting} refers to approaches in which a large language model is provided with task instructions without including exemplar input–output pairs in the prompt context. In this setting, prompts typically specify the modelling objective, target diagram type, and expected output representation (for example, a PlantUML specification), relying on instruction-only guidance to elicit a diagram-related output \cite{xiao2025uml}, \cite{abukhalaf2023codex}, \cite{shehata2024creating}, \cite{hassine2025evaluating}, \cite{xiao2025uml}, \cite{abukhalaf2023codex}, \cite{garaccione_evaluating_2025}, \cite{g_s_comparative_2025}. 

In addition to general instruction prompts, several studies employ instruction-only prompts that embed structured constraints directly in the prompt text. These include fixed prompt templates, explicit output schemas, and enumerations of required diagram elements or notation-specific rules, without providing exemplars \cite{reinhartz2025leveraging}, \cite{de2024evaluating}, \cite{xiao2025uml}, \cite{ferrari2024model}, \cite{abukhalaf2023codex}, \cite{al-ahmad_student-centric_2025}, \cite{fill_conceptual_2023}. Some works also use zero-shot prompting as one configuration among alternative prompting variants to examine differences under controlled prompt formulations \cite{abukhalaf2023codex}, \cite{li2024llm}, \cite{guizzardi_assessing_2025}, \cite{chen2023automated}.

\textbf{2) Few-shot prompting} refers to approaches in which the prompt includes one or more example input–output pairs to illustrate the expected task behavior. In the reviewed studies, few-shot prompting is used primarily in diagram construction and transformation tasks, where example specifications are paired with corresponding class, activity, or other diagram representations \cite{chaaben2023towards}, \cite{ben2024software}, \cite{li2024llm}. Few-shot prompting is also applied in refinement and transformation settings, where exemplar pairs are used to guide the conversion of models into alternative representations \cite{abukhalaf2023codex}, \cite{avignone_exploring_2025}.

\textbf{3) Iterative and interactive prompting} refers to approaches in which a diagram-related task is carried out through multiple prompt–response exchanges instead of a single instruction. The model produces an initial output, and subsequent prompts are used to clarify requirements, correct issues, or extend the artifact. In this way, the final result is developed progressively over several interactions.

In the reviewed studies, this strategy appears in settings where additional information is provided incrementally or where intermediate outputs are refined before producing a final diagram-related artifact. For example, study \cite{shehata2024creating} reports an incremental interaction pattern in source-code-to-diagram transformation, where source code is supplied progressively and diagram elements are generated once sufficient context has been established. Study \cite{wang2404llms} analyses conversational exchanges between users and LLMs in diagram-centric tasks and treats the dialogue process as part of the technical approach.

\subsubsection{RQ3.3: What architectural and workflow patterns are used to integrate LLMs?}\label{RQ3.3}
While the reviewed studies address a variety of diagram-related tasks, they also differ in how LLMs are integrated into modelling workflows. Beyond model selection, researchers adopt different architectural designs, retrieval mechanisms, and workflow structures surrounding the model. To capture these variations, we analyse the studies from a system-level perspective, focusing on how LLMs are embedded within diagram-centric software engineering pipelines. The following subsections describe the main integration patterns observed in the corpus.

\textbf{1) Standalone LLM invocation} refers to approaches in which a diagram-centric task is handled through a single model call. The model receives an input artifact or specification and directly produces the corresponding output.

Several reviewed studies adopt this simple integration pattern in diagram construction and transformation tasks, where textual or structured inputs are mapped directly to diagram representations through one LLM invocation \cite{xiao2025uml}, \cite{abukhalaf2023codex}, \cite{shehata2024creating}, \cite{hassine2025evaluating}, \cite{garaccione_evaluating_2025}, \cite{g_s_comparative_2025}. Similar standalone usage appears in explanation and evaluation contexts, where diagrams are analysed or summarized without intermediate processing stages \cite{li2024llm}, \cite{guizzardi_assessing_2025}.

\textbf{2) Multi-stage LLM workflows} decompose diagram-centric tasks into sequential stages with distinct responsibilities. Instead of processing a task in a single step, these approaches organize modelling activities into explicit processing stages that exchange intermediate artifacts.
Several reviewed studies structure modelling tasks into sequential stages that progressively transform textual inputs into structured diagram representations \cite{wang2025devcoach}, \cite{babaalla2025llm}. A central feature of these workflows is the use of intermediate representations to connect stages. Rather than passing unstructured text between components, these approaches introduce partial diagrams or structured modelling fragments that constrain subsequent processing \cite{ben2024software}, \cite{acher2023generative}, \cite{wang2404llms}.

\textbf{3) Agent-based architectures} refer to approaches in which multiple coordinated components, often implemented as distinct LLM roles, collaborate to complete a diagram-centric task. Rather than structuring processing strictly as a fixed sequence of stages, these systems assign different responsibilities to separate agents, such as interpretation, critique, refinement, or validation.

In the reviewed literature, agent-orchestrated systems adopt explicit role separation, where each agent is assigned a specific modelling function. Study \cite{wang2025devcoach}, for example, decomposes the modelling task across multiple agents with clearly defined responsibilities, enabling diagram generation and refinement through coordinated interactions. More recent work, UCD-LLM \cite{liu_ucd-llm_2026}, introduces a multi-agent framework for use-case diagram modelling in which agents are assigned roles including requirement interpretation, diagram construction, and consistency checking.

\textbf{4) Retrieval-Augmented Generation (RAG)} integrates large language models with external knowledge repositories that are queried at inference time and incorporated into the model’s input. Unlike instruction-only approaches, where relevant context must be included directly in the prompt, RAG architectures retrieve domain-relevant artifacts, and supply them as additional context for downstream processing.

In the reviewed literature, RAG techniques appear in a small studies, primarily in educational support and artifact-linking scenarios. Studies \cite{ardimento2024rag} and \cite{ardimento2024teaching} integrate retrieval mechanisms into UML learning environments, augmenting LLMs with repositories of UML-related artifacts and reference models. The retrieved artifacts are used to ground feedback for students constructing class diagrams, aligning responses with both the current diagram and stored modelling knowledge. LiSSA \cite{fuchss2025lissa} adopts a retrieval-augmented architecture to recover traceability links between heterogeneous software artifacts, including architecture models.

\textbf{5) Human-in-the-Loop Integration(HITL)} integration refers to approaches in which human participation is an explicit and required component of the modelling workflow. Users contribute through requirement specification, clarification, intermediate validation, or revision of artifacts before task completion. 

One common HITL pattern appears in educational support systems, where learners or instructors interact with LLM-based tools to receive guidance, explanations, and corrective feedback during diagram-related activities. Several studies integrate LLMs into tutoring or scaffolding workflows for UML and conceptual modelling, making user interaction central to task progression \cite{ardimento2024enhancing}, \cite{ardimento2024rag}, \cite{ardimento2024teaching}. A second pattern involves interactive clarification and editing during diagram construction. In these systems, users refine requirements or adjust intermediate results as part of the workflow \cite{lu2025aug}.

\subsection{RQ4: What evaluation metrics and datasets are used to assess LLM-based diagram-centric approaches?}\label{subsec2RQ4}

This research question examines how LLM-based approaches for diagram-centric software engineering tasks are evaluated in the reviewed studies. We consider the evaluation metrics and criteria used to assess performance, as well as the datasets and benchmarks employed to support empirical validation. RQ4 is structured into two subquestions:

\indent \begin{itemize}
    \item \textbf{RQ4.1:} What evaluation metrics and criteria are used?
    \item \textbf{RQ4.2:} What datasets and benchmarks support evaluation?
\end{itemize}

\subsubsection{RQ4.1: What evaluation metrics and criteria are used?}\label{RQ4.1}
To answer RQ4.1, we examine the evaluation metrics and assessment criteria reported across the reviewed studies. The literature employs a combination of syntactic validation checks, element-level matching metrics, structural similarity measures, and human-centered assessments. Several of these metrics recur across construction, transformation, and quality assurance contexts, although their operationalization varies depending on artifact type and the availability of reference models.

\indent \begin{itemize}
    \item \textbf{Syntactic Validity:} Syntactic Validity is one of the most frequently reported evaluation criteria, particularly in studies that generate diagrams in executable textual formats such as PlantUML or other serialized UML representations. Generated artifacts are evaluated based on successful parsing, compilation, or validation by downstream tools (e.g., \cite{camara2023assessment}, \cite{de2024evaluating}, \cite{ramachandran2025transforming}, \cite{xiao2025uml}; transformation-oriented validation in \cite{abukhalaf2023codex}, \cite{shehata2024creating}, \cite{abukhalaf2024pathocl}, \cite{moezkarimi2025harnessing}). This criterion ensures well-formedness with respect to the target notation or metamodel but does not guarantee semantic correctness or domain adequacy.
    
    \item\textbf{Precision, Recall, and F1-Score:} When generated or extracted diagram elements can be aligned with a gold-standard reference artifact, studies commonly report Precision, Recall, and F1-score. These element-level matching metrics quantify correctness and completeness of predicted entities, relationships, or model fragments (e.g., \cite{reinhartz2025leveraging}, \cite{tabassum2024using}, \cite{de2024evaluating}, \cite{ahmad2023towards}, \cite{ramachandran2025transforming}, \cite{omar2023measurement}, \cite{xiao2025uml}, \cite{tagliaferro2025leveraging}, \cite{garaccione_evaluating_2025}, \cite{guizzardi_assessing_2025}; structured extraction settings in \cite{fuchss2025lissa}, \cite{hassine2025evaluating}). Their applicability depends on clearly defined matching criteria and the availability of reference diagrams with consistent granularity.

    \item\textbf{Semantic Correctness:} Beyond syntactic validity and element-level matching, many studies assess semantic correctness, evaluating whether a generated or transformed diagram accurately represents the intended domain semantics or source artifact. This evaluation is frequently conducted through manual comparison against expert-authored reference models or structured inspection of element mappings (e.g., \cite{camara2023assessment}, \cite{tabassum2024using}, \cite{de2024evaluating}, \cite{ahmad2023towards}, \cite{babaalla2025llm}, \cite{ramachandran2025transforming}, \cite{omar2023measurement}, \cite{xiao2025uml}, \cite{ferrari2024model}; transformation contexts in \cite{siala2024enhancing}, \cite{moezkarimi2025harnessing}, \cite{abukhalaf2023codex}, \cite{abukhalaf2024pathocl}). In most cases, semantic correctness relies on expert judgment rather than automated semantic oracles.

    \item\textbf{Completeness and Coverage:} Several works measure completeness or requirement coverage, assessing whether required elements, relationships, or constraints are present in the produced diagram (e.g., \cite{zhao2024generating}, \cite{sultan2024ai}, \cite{jahan2024automated}, \cite{babaalla2025llm}, \cite{li2024llm}, \cite{wang2404llms}, \cite{maass_application_2025}, \cite{chen2023automated}). Operationalization varies from counting missing classes or entities to computing requirement-level coverage ratios. These metrics are particularly relevant in requirements-to-model and pipeline-based workflows.

    \item\textbf{Similarity and Distance-Based Measures:} A smaller subset of studies employs structural or semantic similarity measures to quantify deviation between generated and reference artifacts. Reported metrics include Graph Edit Distance (GED) for structural comparison of diagrams (e.g., \cite{de2024evaluating}, \cite{omar2023measurement}), Mean Absolute Percentage Error (MAPE)-style measures for element-level deviation (e.g., \cite{nair_fine-tuned_2025}, \cite{guizzardi_assessing_2025}, \cite{liu_ucd-llm_2026}), and BLEU scores or embedding-based cosine similarity in scenarios involving textual or intermediate representations (e.g., \cite{bates2025unified}, \cite{abukhalaf2023codex}). These metrics aim to approximate structural proximity when direct element-level alignment is difficult, but their adoption remains comparatively limited.

    \item\textbf{Human-Centered Evaluation and Agreement Measures:} Human evaluation plays a central role in several studies, particularly in educational, exploratory, and quality assurance contexts (e.g., \cite{zhao2024generating}, \cite{sultan2024ai}, \cite{jahan2024automated}, \cite{babaalla2025llm}, \cite{li2024llm}, \cite{lu2025aug}, \cite{wang2404llms}, \cite{maass_application_2025}, \cite{fill_conceptual_2023}, \cite{ouh2025evaluating}, \cite{ardimento2024enhancing}, \cite{ardimento2024rag}). Reported methods include expert scoring, rubric-based grading, and Likert-scale assessments of understandability, usefulness, or pedagogical value. In quality assurance settings, studies report agreement between LLM-generated grades and human reference scores (e.g., \cite{ramachandran_ai_2025}, \cite{wang2025assessing}, \cite{ibanez2025can}). However, systematic reporting of inter-rater reliability among human evaluators is inconsistent.
\end{itemize}

\begin{sidewaystable}
\centering
\caption{Evaluation metrics used in LLM-based diagram-centric studies}
\label{tab:rq4_metrics}
\small
\begin{tabular}{l p{4.5cm} p{4.5cm} p{4cm}}
\toprule
\textbf{Metric / Criterion} & 
\textbf{Operational Definition} & 
\textbf{Typical Usage Context} & 
\textbf{Representative Papers} \\
\midrule

Syntactic Validity / Parsability 
& Successful parsing or compilation of generated diagrams 
& Textual diagram generation or model transformation 
& \cite{camara2023assessment}, \cite{de2024evaluating}, \cite{ramachandran2025transforming}, \cite{xiao2025uml}, \cite{abukhalaf2023codex}, \cite{shehata2024creating}, \cite{abukhalaf2024pathocl}, \cite{moezkarimi2025harnessing} \\

\addlinespace
\addlinespace

Precision / Recall / F1-score 
& Element-level matching against reference diagrams 
& Gold-standard alignment available 
& \cite{reinhartz2025leveraging}, \cite{tabassum2024using}, \cite{de2024evaluating}, \cite{ahmad2023towards}, \cite{ramachandran2025transforming}, \cite{omar2023measurement}, \cite{xiao2025uml}, \cite{tagliaferro2025leveraging}, \cite{garaccione_evaluating_2025}, \cite{guizzardi_assessing_2025}, \cite{fuchss2025lissa}, \cite{hassine2025evaluating} \\

\addlinespace
\addlinespace

Semantic Correctness 
& Manual or structured verification of domain fidelity 
& Expert comparison to intended semantics 
& \cite{camara2023assessment}, \cite{tabassum2024using}, \cite{de2024evaluating}, \cite{ahmad2023towards}, \cite{babaalla2025llm}, \cite{ramachandran2025transforming}, \cite{omar2023measurement}, \cite{xiao2025uml}, \cite{ferrari2024model}, \cite{siala2024enhancing}, \cite{moezkarimi2025harnessing}, \cite{abukhalaf2023codex} \\

\addlinespace
\addlinespace

Completeness / Coverage 
& Measurement of missing entities or requirement coverage 
& Requirements-to-model workflows 
& \cite{zhao2024generating}, \cite{sultan2024ai}, \cite{jahan2024automated}, \cite{babaalla2025llm}, \cite{li2024llm}, \cite{wang2404llms}, \cite{maass_application_2025}, \cite{chen2023automated} \\

\addlinespace
\addlinespace

Graph Edit Distance (GED) 
& Structural edit distance between diagram graphs 
& Structural similarity analysis 
& \cite{de2024evaluating}, \cite{omar2023measurement} \\

\addlinespace
\addlinespace

Mean Absolute Percentage Error (MAPE) 
& Proportional deviation of predicted elements 
& Controlled structural evaluation 
& \cite{nair_fine-tuned_2025}, \cite{guizzardi_assessing_2025}, \cite{liu_ucd-llm_2026} \\

\addlinespace
\addlinespace

BLEU / Embedding Similarity 
& Textual or embedding-based similarity comparison 
& Intermediate or textual representations 
& \cite{bates2025unified}, \cite{abukhalaf2023codex} \\

\addlinespace
\addlinespace

Human Evaluation (Rubrics / Likert) 
& Expert scoring or Likert-scale quality assessment 
& Educational and QA settings 
& \cite{zhao2024generating}, \cite{sultan2024ai}, \cite{jahan2024automated}, \cite{babaalla2025llm}, \cite{li2024llm}, \cite{lu2025aug}, \cite{wang2404llms}, \cite{maass_application_2025}, \cite{fill_conceptual_2023}, \cite{ouh2025evaluating}, \cite{ardimento2024enhancing}, \cite{ardimento2024rag} \\

\addlinespace
\addlinespace

Agreement Measures 
& Human–LLM grading consistency 
& Diagram grading or QA tasks 
& \cite{ibanez2025can}, \cite{ramachandran_ai_2025}, \cite{wang2025assessing} \\

\botrule
\end{tabular}
\end{sidewaystable}

\subsubsection{RQ4.2: What datasets and benchmarks support evaluation?}\label{RQ4.2}

To address RQ4.2, we examine the empirical resources used to evaluate LLM-based diagram-centric approaches. Across the reviewed studies, evaluation relies on a combination of publicly available datasets that provide reference diagrams, natural language specifications, and modelling artifacts. Table~\ref{tab:datasets} summarizes the datasets reported in the reviewed literature.

\begin{table*}[t]
\centering
\small
\caption{Public datasets used for evaluating diagram-centric LLM approaches}
\label{tab:datasets}
\begin{tabular}
{p{0.10\linewidth} p{0.35\linewidth} p{0.45\linewidth}}
\toprule
\textbf{Paper} & \textbf{Public URL / DOI} & \textbf{Brief Description} \\
\midrule

\cite{moezkarimi2025harnessing} & \url{https://github.com/gnowin/UML-To-Rebecca-Dataset} & Dataset supporting UML-to-Rebecca model transformation tasks. \\

\cite{abukhalaf2024pathocl} & \url{https://doi.org/10.5281/zenodo.10841785} & UML class models paired with English textual specifications. \\

\cite{shehata2024creating} & \url{http://models-db.com/} & Public repository of software models used for model-driven engineering research. \\

\cite{ardimento2024teaching} & \url{https://modelset.github.io/} & ModelSet repository containing Ecore models for model transformation and analysis tasks. \\

\cite{jahan2024automated} & \url{https://data.mendeley.com/datasets/7zbk8zsd8y/1} & Dataset of software modelling artifacts distributed via Mendeley Data. \\

\cite{jahan2024automated} & \url{https://zenodo.org/records/7529130} & Companion Zenodo dataset supporting evaluation scenarios in the same study. \\

\cite{de2024evaluating} & \url{http://doi.org/10.6084/m9.figshare.25434550} & Collection of natural-language UML class diagram exercises with reference solutions. \\

\cite{antal2024toward} & \url{https://zenodo.org/records/10301347} & Dataset containing UML class diagrams used in diagram generation experiments. \\

\cite{ferrari2024model} & \url{https://doi.org/10.5281/zenodo.10579731} & Requirements corpus compiled from multiple public sources. \\

\cite{bates2025unified} & \url{https://doi.org/10.5281/zenodo.15103682} & Synthetic dataset of UML diagrams generated from PlantUML templates. \\

\cite{camara2023assessment} & \url{https://github.com/atenearesearchgroup/chatgpt-uml} & Dataset and artifacts used for ChatGPT-based UML diagram generation experiments. \\

\cite{ojha_towards_2025} & \url{https://zenodo.org/record/4121935} & Requirements and modelling dataset used in snowballing study. \\

\cite{rahmanian_challenges_2025} & \url{https://github.com/MojdehRahmanian/MMLLMsForERDEvaluation} & Dataset used for evaluating multimodal LLMs on ER diagram generation tasks. \\

\cite{ramachandran_ai_2025} & \url{https://github.com/athulyaanilkumar/ER-diagram-Dataset} & Dataset of entity–relationship diagrams for diagram extraction and evaluation. \\

\cite{wang2404llms} & \url{https://doi.org/10.5281/zenodo.10532600} & Dataset of modelling artifacts used for evaluation of diagram analysis tasks. \\

\cite{nguyen2025automated} & \url{https://doi.org/10.57967/hf/5932} & HuggingFace dataset used for diagram-centric modelling experiments. \\

\bottomrule
\end{tabular}
\end{table*}

\subsection{RQ5: What limitations characterize current LLM-based approaches for diagram-centric software engineering tasks?}\label{subsec2:RQ5}
This research question examines the limitations and threats reported in the reviewed studies. Although many works demonstrate promising results for diagram construction, transformation, evaluation, and educational support, recurring constraints are consistently documented across different diagram types and technical configurations.

\indent \begin{itemize}
    \item \textbf{modelling Accuracy and Semantic Reliability:} Several studies observe that LLM-generated diagrams may contain hallucinated elements, incorrect relationships, incomplete representations, or violations of metamodel constraints, particularly in more complex or multi-step modelling scenarios (e.g., \cite{camara2023assessment}, \cite{tabassum2024using}, \cite{de2024evaluating}, \cite{omar2023measurement}, \cite{xiao2025uml}, \cite{ferrari2024model}, \cite{nair_fine-tuned_2025}, \cite{guizzardi_assessing_2025}, \cite{liu_ucd-llm_2026}). While syntactic validity can often be achieved, semantic correctness and domain alignment remain sensitive to ambiguity in input specifications.
    \item \textbf{Sensitivity to Prompts and Model Configuration:} A number of works note substantial sensitivity to prompt phrasing, few-shot examples, and model selection (e.g., \cite{zhao2024generating}, \cite{tabassum2024using}, \cite{kuchenbuch2025smart}, \cite{de2024evaluating}, \cite{omar2023measurement}, \cite{ferrari2024model}, \cite{nair_fine-tuned_2025}, \cite{guizzardi_assessing_2025}). Performance differences across model versions are occasionally reported, but systematic ablation studies and cross-model comparisons remain limited.
     \item \textbf{Evaluation Design and Robustness:} Evaluation practices frequently rely on small datasets, single-run experiments, or manually curated case studies (e.g., \cite{al2025student}, \cite{ibanez2025can}, \cite{ouh2025evaluating}, \cite{de2024evaluating}, \cite{li2024llm}, \cite{xiao2025uml}, \cite{bouali_toward_2025}, \cite{ramachandran_ai_2025}). Reporting of statistical significance testing, robustness analysis, and inter-rater agreement is inconsistent. In educational and quality assurance contexts, reliance on human judgment introduces subjectivity that is not always formally quantified.   
     \item \textbf{Dataset and Benchmark Limitations:} The reviewed literature reveals limited reuse of shared benchmarks. Many datasets are study-specific or domain-restricted, and few are reused across multiple papers (e.g., \cite{tabassum2024using}, \cite{kuchenbuch2025smart}, \cite{de2024evaluating}, \cite{shehata2024creating}, \cite{siala2025using}, \cite{xiao2025uml}, \cite{maass_application_2025}). As a result, cross-study comparability remains constrained.
     \item \textbf{Reproducibility and Tooling Constraints:} Several studies acknowledge reproducibility challenges stemming from incomplete reporting of prompts, reliance on proprietary APIs, and non-deterministic LLM behavior (e.g., \cite{tabassum2024using}, \cite{kuchenbuch2025smart}, \cite{de2024evaluating}, \cite{shehata2024creating}, \cite{siala2025using}, \cite{xiao2025uml}, \cite{maass_application_2025}). In some cases, replication artifacts are tightly coupled to specific model versions or temporary repositories.
\end{itemize}

\section{Discussion}\label{sec3:Discussion}

\subsection{Diagram Types and Task Coverage}\label{Dis1}
The results clearly show a strong concentration on structural UML modelling, especially class diagrams. Behavioral diagrams appear less frequently, and ER modelling is represented in a smaller subset of studies. 
This imbalance has implications beyond simple coverage gaps: behavioral diagram types such as sequence diagrams and state machines involve temporal ordering and interaction semantics that pose qualitatively different challenges for LLMs than static structural representations.

At the task level, the main emphasis is on creating diagrams from natural language. Transformation tasks form a secondary focus, while evaluation, quality assurance, and educational support applications are not as commonly addressed. It's worth noting that tasks related to model evolution, ensuring consistency across different views, or the incremental refinement of existing diagrams are rarely discussed, despite their significance in practical modelling workflows.

\subsection{LLM Integration in modelling Workflows}\label{Dis2}
Most studies utilize LLMs through prompt-based interactions, where users input natural language requirements or source code, and the model generates serialized diagram specifications. In many cases, the LLM operates as a standalone generator without deeper integration into modelling toolchains.
This standalone usage pattern contrasts with how LLMs are increasingly integrated in code-centric SE tasks, where tool-augmented architectures incorporating compilers, test suites, and static analysers have become common \cite{yang2024swe}. The modelling domain currently lacks equivalent infrastructure for automated feedback, for instance, metamodel validators or constraint checkers that could be invoked programmatically to verify LLM outputs. This limits the potential for closed-loop generation–validation cycles that have proven effective in code generation \cite{chen2022codet}.

Structured configurations are less frequently seen but come with notable advantages. Multi-stage pipelines, hybrid approaches combining LLM outputs with rule-based checks, and retrieval-augmented mechanisms are reported in a smaller subset of studies. 
Among multi-stage approaches, the decomposition of generation into intermediate steps, such as extracting entities before constructing relationships, appears to reduce hallucinated elements in the resulting diagrams, as reported in \cite{babaalla2025llm} and \cite{wang2025devcoach}. Similarly, studies that embed metamodel constraints or output schemas into prompts report improved syntactic validity \cite{reinhartz2025leveraging,  fill_conceptual_2023}, suggesting that structured prompting partially compensates for the absence of external validation tools.

Several studies that incorporate human-in-the-loop interaction report improved output quality compared to fully automated configurations within their respective experimental settings \cite{lu2025aug,ardimento2024enhancing,ardimento2024rag}. However, these comparisons are not conducted under controlled conditions with matched baselines, making it difficult to attribute improvements solely to human involvement. What can be observed is that current LLM capabilities are better suited for assistive roles in modelling workflows, where human modelers retain decision authority over design choices, rather than for fully autonomous diagram generation. This observation aligns with findings in the broader AI-assisted SE literature, where human-AI collaboration consistently outperforms either humans or AI working alone on complex design tasks \cite{vaithilingam2022expectation}.

\subsection{Evaluation Practices and Empirical Infrastructure}\label{Dis3}
The reviewed studies employ a diverse set of evaluation metrics. Syntactic validity checks, precision and recall, F1-score, similarity measures, and human judgment are all reported across different contexts. However, evaluation designs vary considerably in scope and rigor. Several studies rely on small datasets, controlled exercises, or manually curated case studies.

This diversity partly reflects the inherent complexity of evaluating diagram quality, unlike code generation, where execution-based testing provides an objective correctness signal  \cite{austin2021program}, diagram evaluation requires assessing structural validity, semantic fidelity, and domain appropriateness simultaneously, and no single metric captures all these dimensions \cite{camara2023assessment,de2024evaluating}.
A further complication is the lack of consensus on what constitutes a "correct" diagram. Software models are inherently underspecified: multiple valid designs can satisfy the same set of requirements \cite{france2007model}. This makes element-level matching against a single reference solution problematic, as studies that rely exclusively on precision and recall against one gold standard may penalize legitimate design alternatives.

Public datasets are available in a subset of works, but reuse across studies remains limited. 
 Many evaluations are conducted on study-specific corpora, which constrains comparability. Statistical testing and repeated trials are not consistently reported. This contrasts sharply with adjacent fields where shared benchmarks have accelerated progress. In code generation, benchmarks such as HumanEval \cite{chen_evaluating_2021} enabled standardized comparison across models and techniques, contributing to rapid methodological improvement. In NLP, shared benchmarks such as SQuAD \cite{rajpurkar2016squad} and GLUE \cite{wang2018glue} played a similar catalytic role. The diagram-centric modelling community currently lacks an equivalent resource. As a result, empirical evidence is heterogeneous. While individual studies demonstrate promising results, differences in datasets, metrics, and evaluation protocols make systematic comparison across approaches difficult. Establishing shared, multi-diagram-type benchmarks with agreed-upon evaluation protocols would be an important step toward consolidating empirical evidence in this field.

\subsection{ Research Context and Practical Relevance}
The reviewed studies are predominantly situated in academic or educational contexts. A substantial portion of evaluations rely on textbook-style exercises, student-produced artifacts, or controlled laboratory settings \cite{al-ahmad_student-centric_2025, ouh2025evaluating,de2024evaluating,wang2025assessing} while only a small number evaluate their approaches using industrial requirements\cite{kuchenbuch2025smart,xiao2025uml,sultan2024ai}. This has two main implications. First, the input specifications used in most studies are considerably simpler than what practitioners encounter in real projects, as industrial requirements tend to be incomplete, ambiguous, and distributed across multiple documents \cite{handbook2003contract}, and industrial models can contain hundreds of classes with complex dependencies\cite{petre_uml_2013}. 

Second, the majority of reviewed tools operate as standalone scripts or web interfaces producing PlantUML or Mermaid output, with no integration into established modelling platforms such as Eclipse Papyrus or Enterprise Architect. Additionally, all reviewed evaluations are cross-sectional. No study examines how LLM-assisted modelling performs over the course of a project, how generated diagrams evolve through iterative refinement, or how teams incorporate LLM outputs into collaborative workflows. Longitudinal and field studies\cite{stol2018abc} would provide valuable insights into practical utility and adoption barriers.

\section{Implications and Future Directions}\label{Dis5}

\subsection{Improving LLM Reliability}\label{FD1}

Hallucination remains a major problem as it is common for generated diagrams to contain hallucinated features or incorrect relationships that are syntactically correct but lack semantic basis. Recent research into mitigation strategies for hallucinations in LLMs has proposed various approaches including retrieval-augmented grounding, chain of verification, constrained decoding, and multi-agent verification \cite{dhuliawala_chain--verification_2024}, \cite{ji_survey_2023}, \cite{tonmoy_comprehensive_2024}. The adaptation of these strategies to the modelling domain, which requires satisfaction of metamodel constraints as well as semantics, is a major area of research that needs to be explored.

Another area of development that holds promise is the concept of domain-specific fine-tuning. From the literature reviewed, the majority of studies are based on the application of general-purpose models with prompts. However, a minority of studies have explored the application of parameter-efficient fine-tuning techniques such as LoRA\cite{hu2022lora} and QLoRA\cite{dettmers2023qlora} for modelling tasks. The success of domain-specific models in code generation tasks, whereby code-specific LLMs such as CodeLlama and StarCoder outperform general-purpose LLMs\cite{roziere_code_2023}, provides a promising area of development that could improve the accuracy and reduce the hallucinations of UML and ER diagrams. The availability of public modelling datasets, as presented in the findings of the survey (Table~\ref{tab:datasets}), provides a starting point for development.

The non-deterministic nature of the outputs of LLMs poses new challenges in modelling, in which consistency of results across sessions and reproducibility of results are of critical importance. New methods have been proposed in temperature control, deterministic decoding, and self-consistency reasoning in improving the reliability of results in LLMs\cite{wang_self-consistency_2023}. While these methods have generally been studied in reasoning and answering questions, their use in diagram synthesis and model transformation is largely unexplored. Investigating the impact of these methods on the reliability of modelling results is a promising direction for future research, especially in applications in which consistent results are required across multiple development iterations.

The limitation that has consistently been pointed out in the evaluated literature is that, even if syntactic correctness is ensured, there can be violation of metamodel constraints in the generated diagrams. There are formal metamodels that define valid elements and relationships in UML and ER diagrams. However, in most cases, LLM-based methods heavily rely on prompts without explicitly using metamodel information in generation. This could be addressed by metamodel-aware generation, where metamodel information is explicitly used in generation. There are several strategies that could be employed, including modelling schemas in prompts, constrained decoding, and checking models against formal constraints defined in OCL \cite{abukhalaf2024pathocl,shin2021constrained,scholak2021picard}. This could improve the degree of conformity in the generated models with respect to formal semantics supported by modelling languages.

\subsection{Multi-View Consistency and Cross-Diagram Reasoning}

In general, real-world systems are represented through several coordinated diagrams that represent different viewpoints. While structural models represent the architecture of the system, behavioural diagrams show how the system interacts and how workflows are defined. Achieving consistency between these models has been an open issue in model-driven engineering, particularly since any change in one model has implications on related models, hence the integrity of the system. As shown in RQ2, issues of multi-view consistency checking and coordinated diagram modelling have not been adequately addressed in current LLM-based models.

Future research directions include an examination of how LLMs can be leveraged to facilitate cross-diagram reasoning, including detecting inconsistencies between LLMs, propagating changes across related diagrams, and maintaining alignment between structural and behavioural representations. Integrative research directions that leverage LLM reasoning with existing model analysis techniques such as model synchronization and consistency checking \cite{spanoudakis_inconsistency_2001, nuseibeh_leveraging_2002} are promising for providing more trustworthy multi-view modelling systems.

\subsection{Evaluation Standards and Benchmarks}
Progress in this field requires shared evaluation resources that enable systematic comparison across approaches. A priority is the development of a multi-diagram-type benchmark containing natural language specifications of varying complexity paired with expert-validated reference diagrams across structural, behavioural, and conceptual modelling tasks. Critically, such a benchmark should include multiple valid reference solutions per specification, reflecting the inherent design variability in software modelling where multiple correct designs can satisfy the same set of requirements \cite{france2007model}. The catalytic effect of shared benchmarks has been well demonstrated in adjacent fields, notably HumanEval \cite{chen_evaluating_2021} for code generation and GLUE \cite{wang2018glue} for natural language understanding, and a comparable resource for diagram-centric tasks could play a similar role.
Beyond datasets, standardized matching criteria are needed for element-level comparison. Defining common rules for entity matching, relationship alignment, and tolerance for naming variation, analogous to shared task definitions used in evaluation campaigns such as TREC \cite{voorhees1999trec}, would make results comparable across studies and modelling tools. Human evaluation protocols should also adopt established practices from NLP and SE research. Finally, evaluation designs should account for the non-deterministic nature of LLM outputs by reporting results across multiple runs with statistical significance and confidence intervals.

\section{Threats to Validity}\label{ThreatToVal}
In this section, we discuss potential threats to the validity of this survey and the measures taken to mitigate them.

\textbf{Construct validity} concerns whether the study design accurately reflects the intended research scope. The search string was designed to capture literature on large language models and diagram-centric software engineering tasks. However, relevant studies may use alternative terminology that was not included in the query. To mitigate this risk, multiple digital libraries were searched and backward and forward snowballing was performed.

The definition of diagram types, task categories, and technical configurations involves interpretive judgment. Although classification was based on explicit descriptions in the primary studies, some borderline cases required careful interpretation. To reduce ambiguity, inclusion and exclusion criteria were defined prior to screening, and classification decisions were applied consistently across the dataset.

\textbf{Internal validity} relates to potential bias in study selection, data extraction, and synthesis. Study selection followed a predefined protocol, and inclusion and exclusion criteria were applied systematically. Data extraction was conducted using a structured template covering diagram type, task formulation, model configuration, evaluation metrics, datasets, and reported limitations.

Despite these measures, misinterpretation of experimental settings or incomplete reporting in primary studies may affect extracted data. Where necessary, original publications were re-examined to clarify ambiguities.

\textbf{External validity} concerns the generalizability of the findings. The reviewed literature primarily consists of academic studies, often evaluated in controlled or educational settings. As a result, conclusions may not directly generalize to industrial-scale modelling environments.

In addition, the rapid evolution of LLM architectures means that results reported in primary studies are influenced by the specific model versions available at the time of experimentation. Consequently, the landscape described in this survey reflects the state of research within the review period and may evolve as models and evaluation infrastructures mature.

\textbf{Conclusion validity} relates to the reliability of the synthesized findings. The reviewed studies employ heterogeneous evaluation metrics, datasets, and experimental designs, which limits the feasibility of direct quantitative aggregation. The survey therefore focuses on identifying recurring patterns and trends rather than computing aggregated performance measures.

To reduce subjective interpretation, results and discussion are clearly separated, and conclusions are derived from consistent observations across multiple studies rather than isolated findings.

\section{Conclusion}\label{conclusion}
This paper presented a  survey of LLM-based approaches for diagram-centric software engineering tasks. By analyzing diagram types, task formulations, technical integration strategies, evaluation practices, datasets, and reported limitations, the study provides a structured overview of the current research landscape.

The findings indicate a strong concentration on structural UML modelling, particularly class diagrams, with diagram construction from natural language as the dominant task. Evaluation practices are diverse and heterogeneous, and shared benchmark reuse remains limited. While LLMs demonstrate promising capabilities in controlled modelling scenarios, recurring constraints related to semantic reliability, methodological variability, and reproducibility are consistently reported.

Taken together, the results suggest that progress in this area depends not only on advances in model capability, but also on stronger methodological alignment, improved empirical infrastructure, and closer integration with formal modelling semantics. This survey aims to provide a foundation for systematic development and evaluation of future diagram-centric LLM systems within software engineering.






\bibliography{sn-bibliography, Papers}

\end{document}